\documentclass[12pt,prd,onecolumn,showpacs,amsmath,amssymb,aps,floats,floatfix]{revtex4-1}

\usepackage[colorlinks=true,urlcolor=blue,anchorcolor=blue,citecolor=blue,filecolor=blue,linkcolor=blue,menucolor=blue,pagecolor=blue,linktocpage=true]{hyperref} 

\usepackage[inline]{enumitem}
\usepackage[multidot]{grffile}  
\usepackage{dcolumn}
\usepackage{bm}
\usepackage{amsmath}
\usepackage{amsfonts}
\usepackage{amssymb}
\usepackage{color}
\usepackage{latexsym}
\usepackage{slashed} 
\usepackage{pstricks}
\usepackage{indentfirst}
\usepackage{mathrsfs}
\usepackage{multirow}
\usepackage{epsfig,psfrag}
\usepackage{subfigure}
\usepackage{mathtools}
\usepackage{setspace} 
\usepackage[utf8]{inputenc} 
\usepackage[scientific-notation=true]{siunitx} 

\graphicspath{{fig/}}

\newcommand{\SUtwoL}{\mathrm{SU}(2)_\mathrm{L}}
\newcommand{\Uone}{\mathrm{U}(1)}
\newcommand{\svann}{\left<\sigma_\mathrm{ann}v\right>}

\makeatother

\allowdisplaybreaks 

\begin{document}

\title{Pseudo-Nambu-Goldstone dark matter and two-Higgs-doublet models}
\author{Xue-Min Jiang$^{1,2}$}
\author{Chengfeng Cai$^1$}
\author{Zhao-Huan Yu$^1$}\email{yuzhaoh5@mail.sysu.edu.cn}
\author{Yu-Pan Zeng$^1$}
\author{Hong-Hao Zhang$^1$}\email{zhh98@mail.sysu.edu.cn}
\affiliation{$^1$School of Physics, Sun Yat-Sen University, Guangzhou 510275, China}
\affiliation{$^2$Department of Physics, Yunnan University, Kunming 650091, China}

\begin{abstract}

We study a dark matter model with one singlet complex scalar and two Higgs doublets.
The scalar potential respects a softly broken global symmetry, which makes the imaginary part of the singlet become a pseudo-Nambu-Goldstone boson acting as a dark matter candidate.
The pseudo-Nambu-Goldstone nature of the boson leads to the vanishing of its tree-level scattering amplitude off nucleons at zero momentum transfer.
Therefore, although the interaction strength could be sufficiently large to yield a viable relic abundance via thermal mechanism, direct detection is incapable of probing this candidate.
We further investigate the constraints from Higgs measurements, relic abundance observation, and indirect detection.

\end{abstract}

\maketitle
\tableofcontents

\section{Introduction}

Astrophysical and cosmological observations suggest that the majority of matter in the present Universe consists of a nonluminous component called dark matter (DM).
In the conventional paradigm, dark matter is a thermal relic remaining from the early Universe, implying that the interaction strength between DM and standard model (SM) particles may be comparable to the strength of weak interactions~\cite{Bertone:2004pz,Feng:2010gw,Young:2016ala}.
However, null signal results from recent direct detection experiments have put rather stringent constraints on the DM-nucleon scattering cross section~\cite{Akerib:2016vxi,Cui:2017nnn,Aprile:2018dbl}.
This has become a great challenge to the thermal DM paradigm.

A natural way out is to suppress DM-nucleon scattering in direct detection experiments without suppressing DM annihilation in the early Universe.
One possibility is that there are some blind spots with particular parameters leading to the suppression of the DM couplings relevant to direct detection~\cite{Cheung:2012qy,Banerjee:2016hsk,Cai:2017wdu,Han:2018gej,Altmannshofer:2019wjb}.
Additionally, the relevant DM couplings could vanish due to special symmetries~\cite{Dedes:2014hga,Tait:2016qbg,Arcadi:2016kmk,Cai:2016sjz,Xiang:2017yfs,Wang:2017sxx}.
Moreover, DM-nucleon scattering mediated by pseudoscalars can evade direct detection constraints~\cite{Ipek:2014gua,Berlin:2015wwa,No:2015xqa,Goncalves:2016iyg,Haisch:2016gry,Bauer:2017ota,Tunney:2017yfp}.
Furthermore, the DM-nucleon scattering amplitude could be greatly suppressed if the DM particle is a pseudo-Nambu-Goldstone boson (pNGB) protected by an approximate global symmetry~\cite{Barducci:2016fue,Gross:2017dan,Balkin:2018tma,Huitu:2018gbc,Alanne:2018zjm,Kannike:2019wsn,Karamitros:2019ewv,Cline:2019okt}.

In the last case, tree-level interactions of a pNGB are generally momentum suppressed.
As direct detection experiments essentially operate in the zero momentum transfer limit, the amplitude of pNGB dark matter scattering off nucleons vanishes at tree level~\cite{Gross:2017dan}.
Loop corrections could break the global symmetry, resulting in nonvanishing scattering.
Nevertheless, further investigations have shown that the DM-nucleon cross section at one-loop level is rather small, far away from the capability of current direct detection experiments~\cite{Azevedo:2018exj,Ishiwata:2018sdi}.
Therefore, such a pNGB DM framework seems very appealing for thermal DM.

Previous studies in this framework assumed that the Higgs sector just involves one Higgs doublet as in the SM~~\cite{Barducci:2016fue,Gross:2017dan,Balkin:2018tma,Huitu:2018gbc,Alanne:2018zjm,Kannike:2019wsn,Karamitros:2019ewv,Cline:2019okt,Azevedo:2018exj,Ishiwata:2018sdi}.
In this work, we would like to extend the study to two Higgs doublets~\cite{Branco:2011iw}.
A Higgs sector with two $\SUtwoL$ doublets has fairly good motivations.
First, two Higgs doublets are typically required for constructing realistic supersymmetric~\cite{Haber:1984rc} and axion~\cite{Kim:1986ax} models.
Second, the flexible scalar mass spectrum  and additional $CP$ violation sources in two-Higgs-doublet models may be helpful for generating a desired baryon asymmetry of the Universe through the baryogenesis mechanism~\cite{Turok:1990zg}.
Finally, two Higgs doublets could provide an available portal to thermal dark matter with attractive phenomenological features~\cite{Ipek:2014gua,Ko:2015fxa,Berlin:2015wwa,No:2015xqa,Goncalves:2016iyg,Bell:2016ekl,Bauer:2017ota,Tunney:2017yfp,Chang:2017gla,Bell:2017rgi,Dey:2019lyr,Altmannshofer:2019wjb}.

In this paper, we consider that the scalar sector involves two $\SUtwoL$ Higgs doublets as well as a complex scalar $S$, which is a SM gauge singlet.
Most terms in the scalar potential obey a global $\Uone$ symmetry $S\to e^{i\alpha}S$.
The exception is a quadratic term that softly breaks this symmetry and gives mass to the imaginary part of $S$, denoted as $\chi$.
The real scalar $\chi$ is what we call pNGB dark matter.
Its pNGB nature makes its scattering amplitude off nucleons vanish at tree level, evading direct detection constraints.
Nonetheless, it is able to obtain an observed DM relic abundance via the thermal production mechanism.
We will perform a random scan in the parameter space to investigate reasonable parameter points that satisfy current Higgs measurements at the Large Hadron Collider (LHC), observation of the DM relic abundance, and constraints from indirect detection experiments.

The paper is organized as follows. In Sec.~\ref{sec:model}, we describe the details of the pNGB DM model with two Higgs doublets, including the scalar potential, mass eigenstates, four types of Yukawa couplings, the vanishing of the DM-nucleon scattering amplitude, and the alignment limit.
In Sec.~\ref{sec:phen}, we perform a random scan in the parameter space and investigate phenomenological constraints from LHC Higgs measurements, relic abundance observation, and indirect detection.
Section~\ref{sec:concl} gives the conclusions and outlook.
In Appendix~\ref{app:tri_coup}, we write down the scalar and gauge trilinear couplings.
Appendix~\ref{app:width} gives some expressions for decay widths of the SM-like Higgs boson.

\section{Model details}
\label{sec:model}

In this section, we study the model details.
As explained above, we assume that the scalar sector involves two Higgs doublets and one SM gauge singlet, and there is a softly broken global $\Uone$ symmetry leading to pNGB dark matter.
The fermion content is assumed to be the same as in the SM.
Analogous to generic two-Higgs-doublet models, there are four types of Yukawa couplings that do not induce flavor-changing neutral currents (FCNCs) at tree level.
We find that these four types are all applicable to our purpose.

\subsection{Scalar potential}

The two $\SUtwoL$ Higgs-doublet fields are denoted as $\Phi_1$ and $\Phi_2$, both carrying hypercharge $+1/2$.
The complex scalar $S$ is a $\SUtwoL$ singlet and carries no hypercharge.
For simplicity, we make two common assumptions for the scalar potential.
The first assumption is that $CP$ is conserved in the scalar sector, leading to only real coefficients.
The second one is that there is a $Z_2$ symmetry $\Phi_1\to -\Phi_1$ or $\Phi_2\to -\Phi_2$ forbidding quartic terms that are odd in either $\Phi_1$ or $\Phi_2$, but such a symmetry can be softly broken by quadratic terms.

Under these assumptions, the general terms in the scalar potential constructed with $\Phi_1$ and $\Phi_2$ are given by~\cite{Branco:2011iw}
\begin{eqnarray}\label{eq:V1}
{V_1} &=& m_{11}^2|{\Phi _1}|^2 + m_{22}^2|{\Phi _2}|^2 - m_{12}^2(\Phi _1^\dag {\Phi _2} + \Phi _2^\dag {\Phi _1}) + \frac{{{\lambda _1}}}{2}|{\Phi _1}{|^4} + \frac{{{\lambda _2}}}{2}|{\Phi _2}{|^4}
\nonumber\\
&& + {\lambda _3}|{\Phi _1}|^2|{\Phi _2}|^2 + {\lambda _4}|\Phi _1^\dag {\Phi _2}|^2 + \frac{{{\lambda _5}}}{2}[{(\Phi _1^\dag {\Phi _2})^2} + {(\Phi _2^\dag {\Phi _1})^2}].
\end{eqnarray} 
And we can write down the potential terms that involve $S$ and respect a global $\Uone$ symmetry $S\to e^{i\alpha}S$,
\begin{equation}\label{eq:V2}
{V_2} =  - m_S^2|S|^2 + \frac{{{\lambda _S}}}{2}|S{|^4} + {\kappa _1}|{\Phi _1}|^2|S|^2 + {\kappa _2}|{\Phi _2}|^2|S|^2 .
\end{equation}
In addition, we introduce a quadratic term softly breaking the global $\Uone$ symmetry,
\begin{equation}
{V_{{\mathrm{soft}}}} =  - \frac{{m'^2_S}}{4}{S^2} + \mathrm{H.c.}
\end{equation}
Note that even if $m'^2_S$ is complex, we can always make it real and positive by a phase redefinition of $S$.
Then $V_2$ and $V_\mathrm{soft}$ respect a dark $CP$ symmetry $S\to S^*$~\cite{Gross:2017dan,Karamitros:2019ewv}.
The soft breaking term $V_\mathrm{soft}$ can be justified by treating $m'^2_S$ as a spurion, arising from a more fundamental theory that does not induce other soft breaking terms involving odd powers of $S$~\cite{Gross:2017dan,Huitu:2018gbc}.

Now the whole scalar potential is
\begin{equation}\label{eq:V}
V = V_1 + V_2 + V_\mathrm{soft}.
\end{equation}
In particular regions of the parameter space, $\Phi_1$, $\Phi_2$, and $S$ develop nonzero vacuum expectation values (VEVs) $v_1$, $v_2$, and $v_s$.
They can be expanded as
\begin{equation}
{\Phi _1} = \begin{pmatrix}
   {\phi _1^ + }  \\
   {({v_1} + {\rho _1} + i{\eta _1})/\sqrt 2 }  \\
 \end{pmatrix},\quad {\Phi _2} = \begin{pmatrix}
   {\phi _2^ + }  \\
   {({v_2} + {\rho _2} + i{\eta _2})/\sqrt 2 }  \\
 \end{pmatrix},\quad S = \frac{{{v_s} + s + i\chi }}{{\sqrt 2 }}.
\end{equation}
By minimizing the potential, we find the following stationary point conditions:
\begin{eqnarray}
m_{11}^2 &=& \frac{{{v_2}}}{{{v_1}}}{m}_{12}^2 - \frac{1}{2}{\lambda _1}v_1^2 - \frac{1}{2}\lambda_{345}v_2^2 - \frac{1}{2}{\kappa _1}v_s^2, \\
m_{22}^2 &=& \frac{{{v_1}}}{{{v_2}}}{m}_{12}^2 - \frac{1}{2}{\lambda _2}v_2^2 - \frac{1}{2}\lambda_{345}v_1^2 - \frac{1}{2}{\kappa _2}v_s^2, \\
m_S^2 &=&  - \frac{1}{2}m'^2_S + \frac{1}{2}{\lambda _S}v_s^2 + \frac{1}{2}{\kappa _1}v_1^2 + \frac{1}{2}{\kappa _2}v_2^2,
\label{eq:stationary:3}
\end{eqnarray}
where 
\begin{equation}
\lambda_{345} \equiv {\lambda _3} + {\lambda _4} + {\lambda _5}.
\end{equation}

Note that all terms in $V_2$ and $V_\mathrm{soft}$ are products of $|S|^2$ or of ${S^2} + {({S^*})^2}$.
As their expansions are
\begin{equation}
|S|^2 = \frac{1}{2}(v_s^2 + {s^2} + {\chi ^2}) + {v_s}s,\quad {S^2} + {({S^*})^2} = v_s^2 + {s^2} - {\chi ^2} + 2{v_s}s,
\end{equation}
the real scalar $\chi$ always appears in pair in the scalar potential.
Therefore, $\chi$ cannot decay, becoming a stable DM candidate.

\subsection{Mass eigenstates}

After the scalar fields obtain their VEVs, the mass squared of $\chi$ is
\begin{equation}
m_\chi ^2 =  - m_S^2 + \frac{1}{2}m'^2_S + \frac{1}{2}{\lambda _S}v_s^2 + \frac{1}{2}{\kappa _1}v_1^2 + \frac{1}{2}{\kappa _2}v_2^2  = m'^2_S,
\end{equation}
where the terms with VEVs are totally canceled by the third stationary point condition \eqref{eq:stationary:3}.
If $m'^2_S=0$, there is no soft breaking term, and $\chi$ is a massless Nambu-Goldstone boson.
If $m'^2_S>0$, $\chi$ would have a physical mass $m_\chi = m'_S$, behaving as a pseudo-Nambu-Goldstone boson.
This is exactly what we want.

The mass terms for the charged scalars are derived as
\begin{equation}
 - \mathcal{L}_{{\mathrm{mass,}}\phi } = 
\left[ {{m}_{12}^2 - \frac{1}{2}({\lambda _4} + {\lambda _5}){v_1}{v_2}} \right]\begin{pmatrix}
   {\phi _1^ - ,} & {\phi _2^ - }  \\
 \end{pmatrix}\begin{pmatrix}
   {{v_2}/{v_1}} & { - 1}  \\
   { - 1} & {{v_1}/{v_2}}  \\
 \end{pmatrix}\begin{pmatrix}
   {\phi _1^ + }  \\
   {\phi _2^ + }  \\
 \end{pmatrix},
\end{equation}
while those for the $CP$-odd scalars are given by
\begin{equation}
 - \mathcal{L}_{{\mathrm{mass,}}\eta } = 
\frac{1}{2}( {{m}_{12}^2 - {\lambda _5}{v_1}{v_2}} )\begin{pmatrix}
   {{\eta _1,}} & {{\eta _2}}  \\
 \end{pmatrix}\begin{pmatrix}
   {{v_2}/{v_1}} & { - 1}  \\
   { - 1} & {{v_1}/{v_2}}  \\
 \end{pmatrix}\begin{pmatrix}
   {{\eta _1}}  \\
   {{\eta _2}}  \\
 \end{pmatrix}.
\end{equation}
The above mass terms can be diagonalized by rotations
\begin{equation}
\begin{pmatrix}
   {\phi _1^ + }  \\
   {\phi _2^ + }  \\
 \end{pmatrix} = R(\beta )\begin{pmatrix}
   {{G^ + }}  \\
   {{H^ + }}  \\
 \end{pmatrix},\quad
\begin{pmatrix}
   {{\eta _1}}  \\
   {{\eta _2}}  \\
 \end{pmatrix} = R(\beta )\begin{pmatrix}
   {{G^0}}  \\
   a  \\
 \end{pmatrix},\quad
R(\beta ) = \begin{pmatrix}
   {\cos \beta } & { - \sin \beta }  \\
   {\sin \beta } & {\cos \beta }  \\
 \end{pmatrix},
\end{equation}
where the rotation angle $\beta$ satisfies 
\begin{equation}
\tan\beta = \frac{v_2}{v_1}.
\end{equation}
Now $G^\pm$ and $G^0$ are massless Nambu-Goldstone bosons eaten by the weak gauge bosons $W^\pm$ and $Z$, while $H^\pm$ and $a$ are physical states with masses
\begin{equation}\label{eq:mHp_ma}
m_{{H^ + }}^2 = \frac{{v_1^2 + v_2^2}}{{{v_1}{v_2}}}\left[ {{m}_{12}^2 - \frac{1}{2}({\lambda _4} + {\lambda _5}){v_1}{v_2}} \right],\quad
m_a^2 = \frac{{v_1^2 + v_2^2}}{{{v_1}{v_2}}}( {{m}_{12}^2 - {\lambda _5}{v_1}{v_2}} ).
\end{equation}

The $CP$-even scalars $\rho_1$, $\rho_2$, and $s$ mix with each other. Their mass terms are
\begin{equation}
 - {\mathcal{L}_{{\mathrm{mass,}}\rho s}} = \frac{1}{2}\begin{pmatrix}
   {{\rho _1,}} & {{\rho _2,}} & s  \\
 \end{pmatrix}\mathcal{M}_{\rho s}^2\begin{pmatrix}
   {{\rho _1}}  \\
   {{\rho _2}}  \\
   s  \\
 \end{pmatrix},
\end{equation}
where the elements of the $3\times 3$ symmetric mass-squared matrix $\mathcal{M}_{\rho s}^2$ are given by
\begin{eqnarray}
{(\mathcal{M}_{\rho s}^2)_{11}} &=& {\lambda _1}v_1^2 + \frac{{{v_2}}}{{{v_1}}}{m}_{12}^2 ,\quad {(\mathcal{M}_{\rho s}^2)_{22}} = {\lambda _2}v_2^2 + \frac{{{v_1}}}{{{v_2}}}{m}_{12}^2,\quad
{(\mathcal{M}_{\rho s}^2)_{33}} = {\lambda _S}v_s^2,
\\
{(\mathcal{M}_{\rho s}^2)_{12}} &=& \lambda_{345}{v_1}{v_2} - {m}_{12}^2,\quad
{(\mathcal{M}_{\rho s}^2)_{13}} = {\kappa _1}{v_1}{v_s},\quad
 {(\mathcal{M}_{\rho s}^2)_{23}} = {\kappa _2}{v_2}{v_s}.\qquad
\end{eqnarray}
$\mathcal{M}_{\rho s}^2$ can be diagonalized by a $3\times 3$ real orthogonal matrix $O$,
\begin{equation}\label{eq:M_phos:diag}
{O^{\mathrm{T}}}\mathcal{M}_{\rho s}^2O = {\mathrm{diag}}(m_{{h_1}}^2,m_{{h_2}}^2,m_{{h_3}}^2).
\end{equation}
The mass eigenstates $h_i$ ($i=1,2,3$) are then related to the interaction eigenstates by
\begin{equation}
\begin{pmatrix}
   {{\rho _1}}  \\
   {{\rho _2}}  \\
   s  \\
 \end{pmatrix} = O\begin{pmatrix}
   {{h_1}}  \\
   {{h_2}}  \\
   {{h_3}}  \\
 \end{pmatrix}.
\end{equation}
One of $h_i$ should behave like the SM Higgs boson in order to be consistent with observation.
Below we adopt a convention with $m_{h_1} \leq m_{h_2} \leq m_{h_3}$.

From the covariant kinetic terms
\begin{equation}\label{eq:L_kin}
{\mathcal{L}_{{\mathrm{kin}}}} = {({D^\mu }{\Phi _1})^\dag }{D_\mu }{\Phi _1} + {({D^\mu }{\Phi _2})^\dag }{D_\mu }{\Phi _2},
\end{equation}
we derive the mass terms for the weak gauge bosons,
\begin{equation}
{\mathcal{L}_{{\mathrm{mass,}}WZ}} = \frac{{{g^2}}}{4}(v_1^2 + v_2^2){W^{ - ,\mu }}W_\mu ^ +  + \frac{1}{2}\frac{{{g^2}}}{{4c_{\mathrm{W}}^2}}(v_1^2 + v_2^2){Z^\mu }{Z_\mu },
\end{equation}
where $c_\mathrm{W} \equiv \cos\theta_\mathrm{W}$ with $\theta_\mathrm{W}$ denoting the Weinberg angle, and $g$ is the $\SUtwoL$ gauge coupling.
Defining $v \equiv \sqrt {v_1^2 + v_2^2} $, the masses of $W$ and $Z$ bosons become
\begin{equation}
{m_W} = \frac{{gv}}{2},\quad {m_Z} = \frac{{gv}}{{2{c_{\mathrm{W}}}}},
\end{equation}
just as in the SM.
From the Fermi constant ${G_{\mathrm{F}}} = {g^2}/(4\sqrt 2 m_{\mathrm{W}}^2)$, we obtain $v = {(\sqrt 2 {G_{\mathrm{F}}})^{ - 1/2}} = 246.22~\si{GeV}$.
Note that $v_1$ and $v_2$ satisfy $v_1 = v c_\beta$ and $v_2 = v s_\beta$, where we have used the shorthand notations $s_\beta \equiv \sin\beta$ and $c_\beta \equiv \cos\beta$.

The scalar and gauge trilinear couplings of the scalar mass eigenstates can be found in Appendix~\ref{app:tri_coup}.

\subsection{Yukawa couplings}

Unlike the standard model, Yukawa couplings between the two Higgs doublets and SM fermions generally lead to tree-level FCNCs, which could cause phenomenological problems in flavor physics.
This is because diagonalizing the fermion mass matrix cannot make sure that the Yukawa interactions are also diagonalized.
Nevertheless, if all fermions with the same quantum numbers just couple to the one same Higgs doublet, the FCNCs will be absent at tree level~\cite{Glashow:1976nt,Paschos:1976ay,Branco:2011iw,Camargo:2019ukv}.
This can be achieved by assuming particular $Z_2$ symmetries for the Higgs doublets and fermions.

As a result, there are four independent types of Yukawa couplings without tree-level FCNCs, listed as follows.
\begin{eqnarray}
&& \text{Type I:}\quad
{\mathcal{L}_{{\mathrm{Y,I}}}} =  - {y_{{\ell _i}}}{{\bar L}_{i{\mathrm{L}}}}{\ell _{i{\mathrm{R}}}}{\Phi _2} - \tilde y_d^{ij}{{\bar Q}_{i{\mathrm{L}}}}{d'_{j{\mathrm{R}}}}{\Phi _2} - \tilde y_u^{ij}{{\bar Q}_{i{\mathrm{L}}}}{u'_{j{\mathrm{R}}}}{{\tilde \Phi }_2} + \mathrm{H.c.}
\\
&& \text{Type II:}\quad
{\mathcal{L}_{{\mathrm{Y,II}}}} =  - {y_{{\ell _i}}}{{\bar L}_{i{\mathrm{L}}}}{\ell _{i{\mathrm{R}}}}{\Phi _1} - \tilde y_d^{ij}{{\bar Q}_{i{\mathrm{L}}}}{d'_{j{\mathrm{R}}}}{\Phi _1} - \tilde y_u^{ij}{{\bar Q}_{i{\mathrm{L}}}}{u'_{j{\mathrm{R}}}}{{\tilde \Phi }_2} + \mathrm{H.c.}
\\
&& \text{Lepton specific:}\quad
{\mathcal{L}_{{\mathrm{Y,L}}}} =  - {y_{{\ell _i}}}{{\bar L}_{i{\mathrm{L}}}}{\ell _{i{\mathrm{R}}}}{\Phi _1} - \tilde y_d^{ij}{{\bar Q}_{i{\mathrm{L}}}}{d'_{j{\mathrm{R}}}}{\Phi _2} - \tilde y_u^{ij}{{\bar Q}_{i{\mathrm{L}}}}{u'_{j{\mathrm{R}}}}{{\tilde \Phi }_2} + \mathrm{H.c.}
\\
&& \text{Flipped:}\quad
{\mathcal{L}_{{\mathrm{Y,F}}}} =  - {y_{{\ell _i}}}{{\bar L}_{i{\mathrm{L}}}}{\ell _{i{\mathrm{R}}}}{\Phi _2} - \tilde y_d^{ij}{{\bar Q}_{i{\mathrm{L}}}}{d'_{j{\mathrm{R}}}}{\Phi _1} - \tilde y_u^{ij}{{\bar Q}_{i{\mathrm{L}}}}{u'_{j{\mathrm{R}}}}{{\tilde \Phi }_2} + \mathrm{H.c.}
\end{eqnarray}
Here ${\tilde \Phi }_2\equiv i\sigma^2 \Phi_2^*$, $L_{i\mathrm{L}} \equiv (\nu_{i\mathrm{L}}, \ell_{i\mathrm{L}})^\mathrm{T}$,
and $Q_{i\mathrm{L}} \equiv (u'_{i\mathrm{L}}, d'_{i\mathrm{L}})^\mathrm{T}$.
The down-type and up-type quark Yukawa matrices $\tilde y_d^{ij}$ and $\tilde y_u^{ij}$ can be diagonalized through $({U_d})_{ij}^\dag \tilde y_d^{jk}{({U_d})_{kl}} = {y_{{d_i}}}{\delta _{il}}$ and $({U_u})_{ij}^\dag \tilde y_u^{jk}{({U_u})_{kl}} = {y_{{u_i}}}{\delta _{il}}$.
Thus, the interaction eigenstates $u'_i$ and $d'_i$ are related to the mass eigenstates $u_i$ and $d_i$ via ${d'_i} = {({U_d})_{ij}}{d_j}$ and ${u'_i} = {({U_u})_{ij}}{u_j}$.
The Cabibbo-Kobayashi-Maskawa matrix is defined as ${V_{ij}} \equiv ({U_u})_{ik}^\dag {({U_d})_{kj}}$.
As we would not discuss neutrino physics in this work, we assume the lepton sector is the same as in the SM.

After the scalars develop the VEVs, the Yukawa interactions provide mass terms to the fermions.
For the mass eigenstates, the four types of Yukawa terms can be expressed in the same form,
\begin{eqnarray}
{\mathcal{L}_{\mathrm{Y}}} &=&   \sum\limits_{f = {\ell _j},{d_j},{u_j}} {\left[- {m_f}\bar ff - \frac{{{m_f}}}{v}\left(\sum\limits_{i = 1}^3\xi _{{h_i}}^f{h_i}\bar ff + \xi _a^fa\bar fi{\gamma _5}f\right) \right]} 
\nonumber\\
&&  - \frac{{\sqrt 2 }}{v}[{H^ + }(\xi _a^{{\ell _i}}{m_{{\ell _i}}}{{\bar \nu }_i}{P_{\mathrm{R}}}{\ell _i} + \xi _a^{{d_j}}{m_{{d_j}}}{V_{ij}}{{\bar u}_i}{P_{\mathrm{R}}}{d_j} + \xi _a^{{u_i}}{m_{{u_i}}}{V_{ij}}{{\bar u}_i}{P_{\mathrm{L}}}{d_j}) + \mathrm{H.c.}],
\end{eqnarray}
where $P_\mathrm{L}$ and $P_\mathrm{R}$ are the left- and right-handed projection operators, respectively.
The coefficients $\xi_{h_i}^f$ and  $\xi_a^f$ are listed in Table~\ref{tab:xi}.

\begin{table}[!t]
\centering
\setlength\tabcolsep{1em}
\renewcommand{\arraystretch}{1.5}
\caption{Coefficients $\xi_{h_i}^f$ and  $\xi_a^f$ in the four types of Yukawa couplings.}
\label{tab:xi}
\begin{tabular}{ccccc}
\hline
\hline
 & Type I & Type II & Lepton specific & Flipped \\
\hline
$\xi_{h_i}^{\ell_j}$ & $O_{2i}/\sin\beta$ & $O_{1i}/\cos\beta$ & $O_{1i}/\cos\beta$ & $O_{2i}/\sin\beta$ \\
$\xi_{h_i}^{d_j}$ & $O_{2i}/\sin\beta$ & $O_{1i}/\cos\beta$ & $O_{2i}/\sin\beta$ & $O_{1i}/\cos\beta$ \\
$\xi_{h_i}^{u_j}$ & $O_{2i}/\sin\beta$ & $O_{2i}/\sin\beta$ & $O_{2i}/\sin\beta$ & $O_{2i}/\sin\beta$ \\
$\xi_a^{\ell_j}$ & $\cot\beta$ & $-\tan\beta$ & $-\tan\beta$ & $\cot\beta$ \\
$\xi_a^{d_j}$ & $\cot\beta$ & $-\tan\beta$ & $\cot\beta$ & $-\tan\beta$ \\
$\xi_a^{u_j}$ & $-\cot\beta$ & $-\cot\beta$ & $-\cot\beta$ & $-\cot\beta$ \\
\hline
\hline
\end{tabular}
\end{table}

\subsection{Vanishing of the DM-nucleon scattering amplitude}

In this subsection, we verify that the tree-level amplitude of DM scattering off nucleons vanishes at zero momentum transfer.
In our case, DM-nucleon scattering is induced by DM-quark scattering.
Therefore, we just need to prove that the DM-quark scattering amplitude vanishes in the zero momentum transfer limit.

\begin{figure}[!t]
\centering
\includegraphics[width=.26\textwidth]{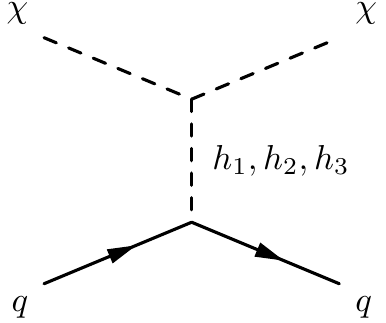}
\caption{Feynman diagram for DM-quark scattering.}
\label{fig:DM-quark}
\end{figure}

From the $U(1)$ symmetric potential~\eqref{eq:V2}, we obtain the trilinear couplings for the DM candidate $\chi$ as
\begin{equation}
{\mathcal{L}_{{\mathrm{tri}},{\chi ^2}}} =  - \frac{1}{2}({\kappa _1}{v_1} {\rho _1} + {\kappa _2}{v_2} {\rho _2} + {\lambda _S}{v_s}s){\chi ^2} = \frac{1}{2}\sum\limits_{i = 1}^3 {{g_{{h_i}{\chi ^2}}}\,{h_i}{\chi ^2}},
\end{equation}
where the coupling coefficients for the mass eigenstates are given by
\begin{equation}\label{eq:g_hchichi}
{g_{{h_i}{\chi ^2}}} =  - {\kappa _1}{v_1} {O_{1i}} - {\kappa _2}{v_2} {O_{2i}} - {\lambda _S}{v_s}{O_{3i}}.
\end{equation}
At tree level, only the $CP$-even Higgs bosons $h_1$, $h_2$, and $h_3$ can mediate $\chi$ scattering off quarks.
The Feynman diagram is shown in Fig.~\ref{fig:DM-quark}.

Take the type-I Yukawa couplings as an example.
Defining a Lorentz invariant $t\equiv p_\mu p^\mu$, where $p^\mu$ is the 4-momentum of the mediator $h_i$, we can write down the DM-quark scattering amplitude as
\begin{equation}
i\mathcal{M} = \frac{{{m_q}}}{{v{s_\beta }}}\bar u({k_2})u({k_1})\left( {{g_{{h_1}{\chi ^2}}}\frac{i}{{t - m_{{h_1}}^2}}{O_{21}} + {g_{{h_2}{\chi ^2}}}\frac{i}{{t - m_{{h_2}}^2}}{O_{22}} + {g_{{h_3}{\chi ^2}}}\frac{i}{{t - m_{{h_3}}^2}}{O_{23}}} \right),
\end{equation}
where $u({k_1})$ and $\bar u({k_2})$ are the wave functions for the incoming and outgoing quarks, respectively.
In the zero momentum transfer limit, $t\to 0$, and the above amplitude can be reexpressed as
\begin{equation}
i\mathcal{M} \to  - i\frac{{{m_q}}}{{v{s_\beta }}}\bar u({k_2})u({k_1})\begin{pmatrix}
   {{g_{{h_1}{\chi ^2}}},} & {{g_{{h_2}{\chi ^2}}},} & {{g_{{h_3}{\chi ^2}}}}  \\
 \end{pmatrix}{(\mathcal{M}_h^2)^{ - 1}}O^\mathrm{T}\begin{pmatrix}
   0  \\
   1  \\
   0  \\
 \end{pmatrix},
\end{equation}
where $(\mathcal{M}_h^2)^{ - 1} = \mathrm{diag}(m_{{h_1}}^{-2},m_{{h_2}}^{-2},m_{{h_3}}^{-2})$ is the inverse of the diagonalized mass-squared matrix $\mathcal{M}_h^2 \equiv \mathrm{diag}(m_{{h_1}}^2,m_{{h_2}}^2,m_{{h_3}}^2)$.
From Eqs.~\eqref{eq:g_hchichi} and \eqref{eq:M_phos:diag}, we have
\begin{equation}
\begin{pmatrix}
   {{g_{{h_1}{\chi ^2}}},} & {{g_{{h_2}{\chi ^2}}},} & {{g_{{h_3}{\chi ^2}}}}  \\
 \end{pmatrix} = -\begin{pmatrix}
   {{\kappa _1}{v_1} ,} & {{\kappa _2}{v_2} ,} & {{\lambda _S}{v_s}}  \\
 \end{pmatrix}O,
\quad
{(\mathcal{M}_h^2)^{ - 1}} = {O^{\mathrm{T}}}{(\mathcal{M}_{\rho s}^2)^{ - 1}}O.
\end{equation}
Utilizing these equations as well as the orthogonality of $O$, we obtain
\begin{equation}
i\mathcal{M} \to i\frac{{{m_q}}}{{v{s_\beta }}}\bar u({k_2})u({k_1})\begin{pmatrix}
   {\kappa _1}{v_1} , & {\kappa _2}{v_2} , & {\lambda _S}{v_s} \\
 \end{pmatrix}{(\mathcal{M}_{\rho s}^2)^{ - 1}}\begin{pmatrix}
   0  \\
   1  \\
   0  \\
 \end{pmatrix}.
\end{equation}
This can be understood as the amplitude expressed in the interaction basis~\cite{Gross:2017dan}.

The inverse of $\mathcal{M}_{\rho s}^2$ can be expressed as its adjugate $\mathcal{A}$ divided by its determinant, i.e., ${(\mathcal{M}_{\rho s}^2)^{ - 1}} = \mathcal{A}/\det (\mathcal{M}_{\rho s}^2)$.
The relevant elements of $\mathcal{A}$ are
\begin{eqnarray}
{\mathcal{A}_{12}} &=&  - ({\lambda _{345}}{v_1}{v_2} - m_{12}^2){\lambda _S}v_s^2 + {\kappa _1}{\kappa _2}{v_1}{v_2} v_s^2,
\\
{\mathcal{A}_{22}} &=& ({\lambda _1}v_1^2 + m_{12}^2\tan \beta ){\lambda _S}v_s^2 - {{\kappa _1^2}{v_1^2} }v_s^2,
\\
{\mathcal{A}_{32}} &=&  - ({\lambda _1}v_1^2 + m_{12}^2\tan \beta ){\kappa _2}{v_2} {v_s} + ({\lambda _{345}}{v_1}{v_2} - m_{12}^2){\kappa _1}{v_1} {v_s}.
\end{eqnarray}
We then have
\begin{equation}
\begin{pmatrix}
   {{\kappa _1}{v_1} }, & {{\kappa _2}{v_2} }, & {{\lambda _S}{v_s}}  \\
 \end{pmatrix}{(\mathcal{M}_{\rho s}^2)^{ - 1}}\begin{pmatrix}
   0  \\
   1  \\
   0  \\
 \end{pmatrix}
= {\det} ^{-1}(\mathcal{M}_{\rho s}^2)({\kappa _1}{v_1} {\mathcal{A}_{12}} + {\kappa _2}{v_2}{\mathcal{A}_{22}} + {\lambda _S}{v_s}{\mathcal{A}_{32}}) = 0.
\end{equation}
Therefore, we have proven that the tree-level DM-quark amplitude $i\mathcal{M}$ vanishes in the zero momentum transfer limit for the type-I Yukawa couplings.
Similarly, we can prove this for the type-II, lepton-specific, and flipped Yukawa couplings.

As the global $\Uone$ symmetry is softly broken, loop corrections would give a nonvanishing DM-nucleon scattering cross section~\cite{Gross:2017dan}.
Nonetheless, we expect that the loop-induced cross section should be typically $\lesssim \mathcal{O}(10^{-50})~\si{cm^2}$, as suggested by the one-loop evaluation in Ref.~\cite{Azevedo:2018exj} where only one Higgs doublet is considered.
Thus, current and near future direct detection experiments should not be able to probe our pNGB DM model.

\subsection{Alignment limit}

Current LHC Higgs measurements favor a $125~\si{GeV}$ SM-like Higgs boson.
If one of the $CP$-even Higgs bosons mimics the SM Higgs boson, the constraints from Higgs measurements can be easily satisfied.
For the two Higgs doublets, such a situation can be achieved by requiring that the additional scalars are much heavier than the weak scale so that the lightest $CP$-even Higgs boson reproduces SM-like Higgs signals at the LHC.
This is known as the ``decoupling'' limit~\cite{Gunion:2002zf}.
In general, a particular parameter set or relation leading to a $CP$-even Higgs boson mimicking the SM Higgs boson is referred as an ``alignment'' limit.
The decoupling limit is of course an alignment limit, but it is less interesting, as the new particles might be too heavy to be accessed at the LHC.

A more interesting possibility is alignment without decoupling~\cite{Carena:2013ooa,Dev:2014yca}.
In order to find such a possibility, we may rotate the two Higgs doublets $\Phi_1$ and $\Phi_2$ into the Higgs basis~\cite{Georgi:1978ri,Donoghue:1978cj}
\begin{equation}
\begin{pmatrix}
   {{\Phi _h}}  \\
   {{\Phi _H}}  \\
 \end{pmatrix} \equiv {R^{ - 1}}(\beta )\begin{pmatrix}
   {{\Phi _1}}  \\
   {{\Phi _2}}  \\
 \end{pmatrix},
\end{equation}
and have
\begin{equation}
{\Phi _h} = \begin{pmatrix}
   {{G^ + }}  \\
   {(v + h + i{G^0})/\sqrt 2 }  \\
 \end{pmatrix},\quad
{\Phi _H}  = \begin{pmatrix}
   {{H^ + }}  \\
   {(H + ia)/\sqrt 2 }  \\
 \end{pmatrix}.
\end{equation}
Now $\Phi_h$ gains a VEV $v$ and contains a $CP$-even scalar $h$ as well as the Nambu-Goldstone bosons, while $\Phi_H$ has zero VEV and contains a $CP$-even scalar $H$ and the physical states $H^+$ and $a$.
Consequently, the tree-level interactions of the $CP$-even scalar $h$ with weak gauge bosons and SM fermions are totally identical to those of the Higgs boson in the SM.
Therefore, the alignment limit means that $h$ does not mix with $H$ and $s$.

In the Higgs basis, the potential terms~\eqref{eq:V1} transform to
\begin{eqnarray}
{V_1}  &=& m_{hh}^2|{\Phi _h}|^2 + m_{HH}^2|{\Phi _H}|^2 - m_{hH}^2(\Phi _h^\dag {\Phi _H} + \Phi _H^\dag {\Phi _h}) + \frac{{{\lambda _h}}}{2}|{\Phi _h}{|^4} + \frac{{{\lambda _H}}}{2}|{\Phi _H}{|^4}
+ {{\tilde \lambda }_3}|{\Phi _h}|^2|{\Phi _H}|^2
\nonumber\\
&&  + {{\tilde \lambda }_4}|\Phi _h^\dag {\Phi _H}|^2 + \frac{1}{2}[{{\tilde \lambda }_5}{(\Phi _h^\dag {\Phi _H})^2} + {{\tilde \lambda }_6}|{\Phi _h}|^2\Phi _H^\dag {\Phi _h} + {{\tilde \lambda }_7}|{\Phi _H}|^2\Phi _h^\dag {\Phi _H} + \mathrm{H.c.}],
\end{eqnarray}
where the new parameters are related to the previous parameters by~\cite{Bell:2017rgi}
\begin{eqnarray}
m_{hh}^2 &=& c_\beta ^2m_{11}^2 + s_\beta ^2m_{22}^2 - 2{s_\beta }{c_\beta }m_{12}^2,\quad m_{HH}^2 = s_\beta ^2m_{11}^2 + c_\beta ^2m_{22}^2 + 2{s_\beta }{c_\beta }m_{12}^2,
\\
m_{hH}^2 &=& {s_\beta }{c_\beta }(m_{11}^2 - m_{22}^2) + (c_\beta ^2 - s_\beta ^2)m_{12}^2,\quad {\lambda _h} = c_\beta ^4{\lambda _1} + s_\beta ^4{\lambda _2} + 2s_\beta ^2c_\beta ^2{\lambda _{345}},
\\
{\lambda _H} &=& s_\beta ^4{\lambda _1} + c_\beta ^4{\lambda _2} + 2s_\beta ^2c_\beta ^2{\lambda _{345}},\quad {{\tilde \lambda }_3} = s_\beta ^2c_\beta ^2({\lambda _1} + {\lambda _2} - 2{\lambda _4} - 2{\lambda _5}) + (s_\beta ^4 + c_\beta ^4){\lambda _3},
\\
{{\tilde \lambda }_4} &=& s_\beta ^2c_\beta ^2({\lambda _1} + {\lambda _2} - 2{\lambda _3} - 2{\lambda _5}) + (s_\beta ^4 + c_\beta ^4){\lambda _4},
\\
{{\tilde \lambda }_5} &=& s_\beta ^2c_\beta ^2({\lambda _1} + {\lambda _2} - 2{\lambda _3} - 2{\lambda _4}) + (s_\beta ^4 + c_\beta ^4){\lambda _5},
\\
{{\tilde \lambda }_6} &=&  - {s_{2\beta }}(c_\beta ^2{\lambda _1} - s_\beta ^2{\lambda _2}) + {s_{2\beta }}{c_{2\beta }}{\lambda _{345}},\quad {{\tilde \lambda }_7} =  - {s_{2\beta }}(s_\beta ^2{\lambda _1} - c_\beta ^2{\lambda _2}) - {s_{2\beta }}{c_{2\beta }}{\lambda _{345}}.
\end{eqnarray}
On the other hand, the potential terms~\eqref{eq:V2} transform to
\begin{equation}
{V_2} =  - m_S^2|S|^2 + \frac{{{\lambda _S}}}{2}|S{|^4} + {{\tilde \kappa }_1}|{\Phi _h}|^2|S|^2 + {{\tilde \kappa }_2}|{\Phi _H}|^2|S|^2 + {{\tilde \kappa }_3}(\Phi _h^\dag {\Phi _H} + \Phi _H^\dag {\Phi _h})|S|^2,
\end{equation}
where the new parameters are given by
\begin{equation}
{{\tilde \kappa }_1} = c_\beta ^2{\kappa _1} + s_\beta ^2{\kappa _2},\quad
{{\tilde \kappa }_2} = s_\beta ^2{\kappa _1} + c_\beta ^2{\kappa _2},\quad
 {{\tilde \kappa }_3} =  - {s_\beta }{c_\beta }({\kappa _1} - {\kappa _2}) .
\end{equation}

Then the stationary point conditions for the scalar potential are
\begin{equation}
m_{hh}^2 =  - \frac{1}{2}{\lambda _h}{v^2} - \frac{1}{2}{{\tilde \kappa }_1}v_s^2,\quad m_{hH}^2 = \frac{1}{4}{{\tilde \lambda }_6}{v^2} + \frac{1}{2}{{\tilde \kappa }_3}v_s^2,\quad m_S^2 =  - \frac{1}{2}m'^2_S + \frac{1}{2}{\lambda _S}v_s^2 + \frac{1}{2}{{\tilde \kappa }_1}{v^2}.
\end{equation}
As a result, the mass-squared matrix for $CP$-even scalars $(h, H, s)$ is
\begin{equation}\setlength\arraycolsep{.5em}
\mathcal{M}_{hHs}^2 = \begin{pmatrix}
   {{\lambda _h}{v^2}} & {{\tilde \lambda }_6}{v^2}/2 & {{{\tilde \kappa }_1}v{v_s}}  \\
   {{\tilde \lambda }_6}{v^2}/2 & {m_{HH}^2 + ({{\tilde \lambda }_{345}}{v^2} + {{\tilde \kappa }_2}v_s^2)/2} & {{{\tilde \kappa }_3}v{v_s}}  \\
   {{{\tilde \kappa }_1}v{v_s}} & {{{\tilde \kappa }_3}v{v_s}} & {{\lambda _S}v_s^2}  \\
 \end{pmatrix}.
\end{equation}
In order to prevent $h$-$H$ and $h$-$s$ mixings, the off-diagonal terms $(\mathcal{M}_{hHs}^2)_{12}$ and $(\mathcal{M}_{hHs}^2)_{13}$ should be absent, corresponding to
\begin{equation}
{{\tilde \lambda }_6} = {{\tilde \kappa }_1} = 0.
\end{equation}
This is the alignment condition in our model.
When this condition is satisfied, the tree-level couplings of $h$ to SM particles are exactly the same as those of the SM Higgs boson.

\section{Phenomenological constraints}
\label{sec:phen}

In this section, we take the type-I Yukawa couplings as an illuminating example to investigate the phenomenological constraints from Higgs measurements, relic abundance observation, and indirect detection.

\subsection{Parameter scan and Higgs measurements}

There are 12 free parameters in the model, which can be chosen as
\begin{equation}
v_s,~~ m_\chi,~~ m_{12}^2,~~ \tan\beta,~~ \lambda_1,~~ \lambda_2,~~ \lambda_3,~~ \lambda_4,~~ \lambda_5,~~ \lambda_S,~~ \kappa_1,~~ \kappa_2.
\end{equation}
In order to investigate the vast parameter space, we carry out a random scan within the following ranges:
\begin{eqnarray}
&& 10~\si{GeV} < v_s < 10^3~\si{GeV},\quad
10~\si{GeV} < m_\chi < 10^4~\si{GeV},
\\
&& (10~\si{GeV})^2 < |m_{12}^2| < (10^3~\si{GeV})^2,\quad
 10^{-2} < \tan\beta < 10^2,
\\
&& 10^{-3} < \lambda_1, \lambda_2, \lambda_S < 1,\quad
10^{-3} < |\lambda_3|, |\lambda_4|, |\lambda_5|, |\kappa_1|, |\kappa_2| < 1.
\end{eqnarray}
Then we require the selected parameter points must give positive $m_{h_{1,2,3}}^2$, $m_{H^+}^2$, and $m_a^2$, ensuring physical scalar masses.
Moreover, one of the $CP$-even Higgs bosons $h_i$ should have a mass within the $3\sigma$ range of the measured SM-like Higgs boson mass $m_h = 125.18\pm 0.16~\si{GeV}$~\cite{Tanabashi:2018oca}.
We recognize this scalar as the SM-like Higgs boson and denote it as $h_\mathrm{SM}$, and further examine if its properties are consistent with current measurements.

In the $\kappa$ framework~\cite{Heinemeyer:2013tqa}, the couplings of the SM-like Higgs boson to SM particles can be expressed as
\begin{eqnarray}
{\mathcal{L}_{{h_{{\mathrm{SM}}}}}} &=& {\kappa _W}g{m_W}{h_{{\mathrm{SM}}}}W_\mu ^ + {W^{ - ,\mu }} + {\kappa _Z}\frac{{g{m_Z}}}{{2{c_{\mathrm{W}}}}}{h_{{\mathrm{SM}}}}{Z_\mu }{Z^\mu } - \sum\limits_f {{\kappa _f}\frac{{{m_f}}}{v}{h_{{\mathrm{SM}}}}\bar ff} 
\nonumber\\
&& + {\kappa _g}g_{hgg}^{{\mathrm{SM}}}{h_{{\mathrm{SM}}}}G_{\mu \nu }^a{G^{a\mu \nu }} + {\kappa _\gamma }g_{h\gamma \gamma }^{{\mathrm{SM}}}{h_{{\mathrm{SM}}}}{A_{\mu \nu }}{A^{\mu \nu }} + {\kappa _{Z\gamma }}g_{Z\gamma \gamma }^{{\mathrm{SM}}}{h_{{\mathrm{SM}}}}{A_{\mu \nu }}{Z^{\mu \nu }},
\end{eqnarray}
where $g_{hgg}^{{\mathrm{SM}}}$, $g_{h\gamma \gamma }^{{\mathrm{SM}}}$, and $g_{Z\gamma \gamma }^{{\mathrm{SM}}}$ are the loop-induced effective couplings to $gg$, $\gamma\gamma$, and $Z\gamma$, respectively.
$\kappa$'s are coupling modifiers, whose values are all equal to 1 in the SM.
Equation~\eqref{eq:g_hiVV} implies that $\kappa _W$ and $\kappa _Z$ are equal in our model, and we will use $\kappa_V$ representing both of them.
Assuming the SM-like Higgs boson is $h_\mathrm{SM} = h_i$, we have
\begin{equation}\label{eq:kapV}
{\kappa _V} =  {c_\beta }{O_{1i}} + {s_\beta }{O_{2i}}.
\end{equation}
The coupling modifiers for fermions can be read off from Table~\ref{tab:xi}.
For the type-I Yukawa couplings, all SM fermions have the same coupling modifier, given by
\begin{equation}\label{eq:kapf}
{\kappa _f} = \frac{{{O_{2i}}}}{{{s_\beta }}}.
\end{equation} 

It is also helpful to define another modifier $\kappa_H$ as
\begin{equation}
\kappa _H^2 \equiv \frac{{{\Gamma _{{h_{{\mathrm{SM}}}}}} - \Gamma _{{h_{{\mathrm{SM}}}}}^{{\mathrm{BSM}}}}}{{\Gamma _h^{{\mathrm{SM}}}}},
\end{equation}
where ${\Gamma _h^{{\mathrm{SM}}}}$ is the Higgs total decay width in the SM, ${{\Gamma _{{h_{{\mathrm{SM}}}}}}}$ is the total decay width of the SM-like Higgs boson $h_\mathrm{SM}$, and $\Gamma _{h_\mathrm{SM}}^\mathrm{BSM}$ is the $h_\mathrm{SM}$ decay width into final states beyond the SM (BSM).
Thus, $\kappa_H$ indicates the deviation of the Higgs width decaying into SM final states and is also equal to 1 in the SM.
In our model, $\Gamma _{h_\mathrm{SM}}^\mathrm{BSM}$ can be generally separated into two parts,
\begin{equation}
\Gamma _{{h_{{\mathrm{SM}}}}}^{{\mathrm{BSM}}} = \Gamma _{{h_{{\mathrm{SM}}}}}^{{\mathrm{inv}}} + \Gamma _{{h_{{\mathrm{SM}}}}}^{{\mathrm{und}}}.
\end{equation}
$\Gamma _{{h_{{\mathrm{SM}}}}}^{{\mathrm{inv}}}$ is the $h_\mathrm{SM}$ decay width into the invisible final state, i.e., a pair of the DM candidate $\chi$.
$\Gamma _{{h_{{\mathrm{SM}}}}}^{{\mathrm{und}}}$ involves decay widths into all kinematically allowed BSM final states that are undetected in current LHC searches.
Such final states may include $aa$, ${H^ + }{H^ - }$, ${h_i}{h_j}$, $aZ$, and $H^\pm W^\mp$.
The expressions for these decay widths are listed in Appendix~\ref{app:width}.
Once all the decay widths are evaluated, we can determine the invisible and undetected BSM branching ratios via $\mathrm{BR}_\mathrm{inv} = \Gamma _{{h_{{\mathrm{SM}}}}}^{{\mathrm{inv}}}/{{\Gamma _{{h_{{\mathrm{SM}}}}}}}$ and $\mathrm{BR}_\mathrm{und} = \Gamma _{{h_{{\mathrm{SM}}}}}^{{\mathrm{und}}}/{{\Gamma _{{h_{{\mathrm{SM}}}}}}}$, respectively.

We utilize a numerical tool \texttt{Lilith 1.1.4}~\cite{Bernon:2015hsa} to study the constraints from current Higgs measurements.
\texttt{Lilith} is able to construct an approximate likelihood based on experimental results of Higgs signal strength measurements.
For each selected parameter point in our random scan, we put the corresponding $m_{h_\mathrm{SM}}$, $\kappa_V$, $\kappa_f$, $\mathrm{BR}_\mathrm{inv}$, and $\mathrm{BR}_\mathrm{und}$ into \texttt{Lilith}.
Then \texttt{Lilith} can evaluate $\kappa_g$, $\kappa_\gamma$, and $\kappa_{Z\gamma}$ involving the loop contributions from SM fermions and gauge bosons whose couplings are modified by $\kappa_f$ and $\kappa_V$, including next-to-leading-order QCD corrections.
Such an evaluation has neglected the loop contributions from the BSM scalars in our model.
Nonetheless, these scalars are typically heavy and/or have small couplings.
Therefore, their contributions are insignificant for most of the selected parameter points.

We further use \texttt{Lilith} to calculate the likelihood $-2\ln L$ for each parameter point based on Tevatron data~\cite{Aaltonen:2013ioz}, ATLAS run~1 data~\cite{Aad:2014iia,Aad:2014eha,Aad:2014xzb,ATLAS:2015yda,Aad:2015gra,Aad:2015iha,Aad:2015ona,Aad:2015gba}, CMS run~1 data~\cite{Chatrchyan:2014tja,Khachatryan:2014qaa,Khachatryan:2014jba,Khachatryan:2015ila,Khachatryan:2015bnx}, ATLAS run~2 data~\cite{ATLAS:2016bza,ATLAS:2016ldo,ATLAS:2016lgh,ATLAS:2016nke,ATLAS:2016oum,ATLAS:2016awy,ATLAS:2016pkl,ATLAS:2016gld,ATLAS:2017syx}, and CMS run~2 data~\cite{CMS:2016nfx,CMS:2016mmc,CMS:2016jjx,CMS:2016ixj,CMS:2016zbb,CMS:2017jkd,CMS:2017lgc}.
We then transform $-2\ln L$ to a $p$-value and require that the selected parameter points should give $p$-values larger than 0.05.
This means that we have rejected the parameter points that are excluded by data at $95\%$ confidence level (C.L.).

\begin{figure}[!t]
\centering
\subfigure[~$\tan\beta$-$\lambda_1$ plane.\label{fig:tanb-lam1}]
{\includegraphics[width=0.48\textwidth]{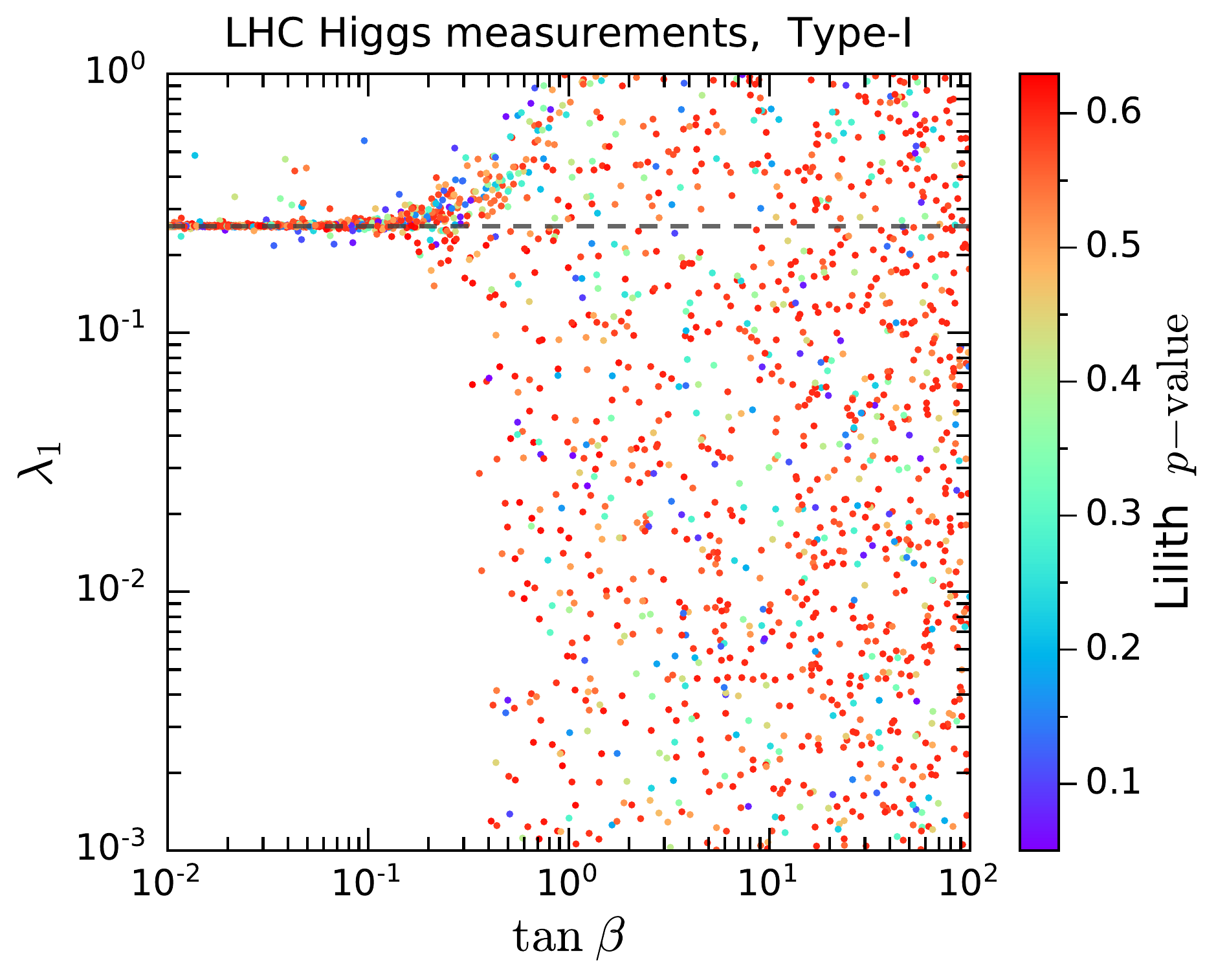}}
\subfigure[~$\tan\beta$-$\lambda_2$ plane.\label{fig:tanb-lam2}]
{\includegraphics[width=0.48\textwidth]{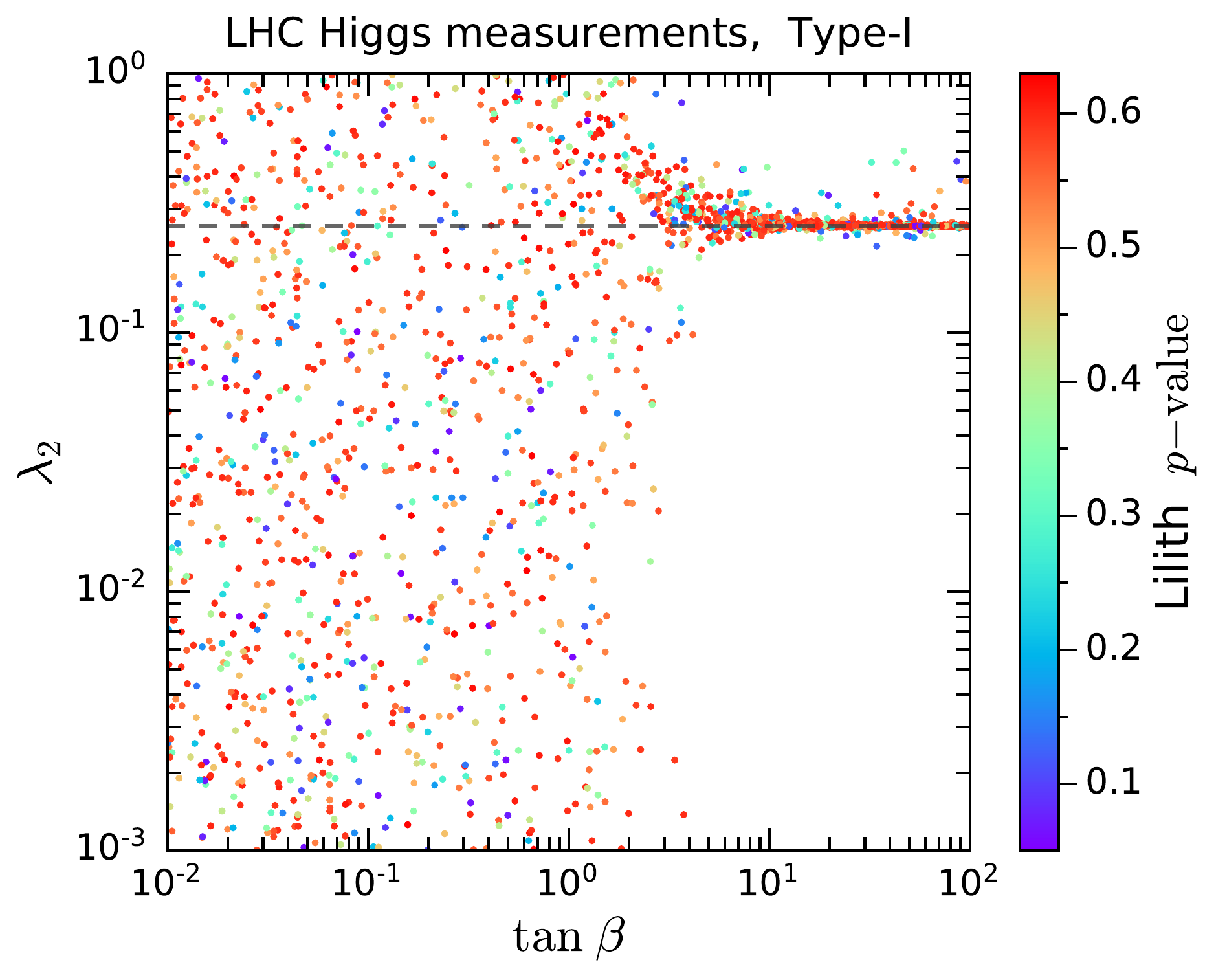}}
\caption{\texttt{Lilith} $p$-values for the selected parameter points projected in the $\tan\beta$-$\lambda_1$ (a) and $\tan\beta$-$\lambda_2$ (b) planes.
The dashed lines indicate the value of $\lambda_\mathrm{SM}$.}
\label{fig:tanb-lam1_lam2}
\end{figure}

Now we can analyze the properties of the remaining parameter points.
Figure~\ref{fig:tanb-lam1_lam2} shows the \texttt{Lilith} $p$-values of the selected parameter points projected in the $\tan\beta$-$\lambda_1$ and $\tan\beta$-$\lambda_2$ planes.
We find that when $\tan\beta\lesssim 0.2$ ($\tan\beta\gtrsim 5$), $\lambda_1$ ($\lambda_2$) tends to converge on $\lambda_\mathrm{SM}=m_h^2/v^2 \simeq 0.26$, which is the quartic Higgs coupling in the SM.
This is because $\tan\beta\ll 1$ ($\tan\beta\gg 1$) leads to $v_1 \gg v_2$ ($v_2 \gg v_1$) and $\Phi_1\simeq \Phi_h$ ($\Phi_2\simeq \Phi_h$), i.e., $\Phi_1$ ($\Phi_2$) acting as the SM-like Higgs doublet.
Since experimental data favor a SM-like Higgs boson, the corresponding quartic coupling would be close to its SM counterpart.

\begin{figure}[!t]
\centering
\subfigure[~$m_{h_\mathrm{SM}}$-$m_{h_1}$ plane.\label{fig:mhSM-mh1}]
{\includegraphics[width=0.48\textwidth]{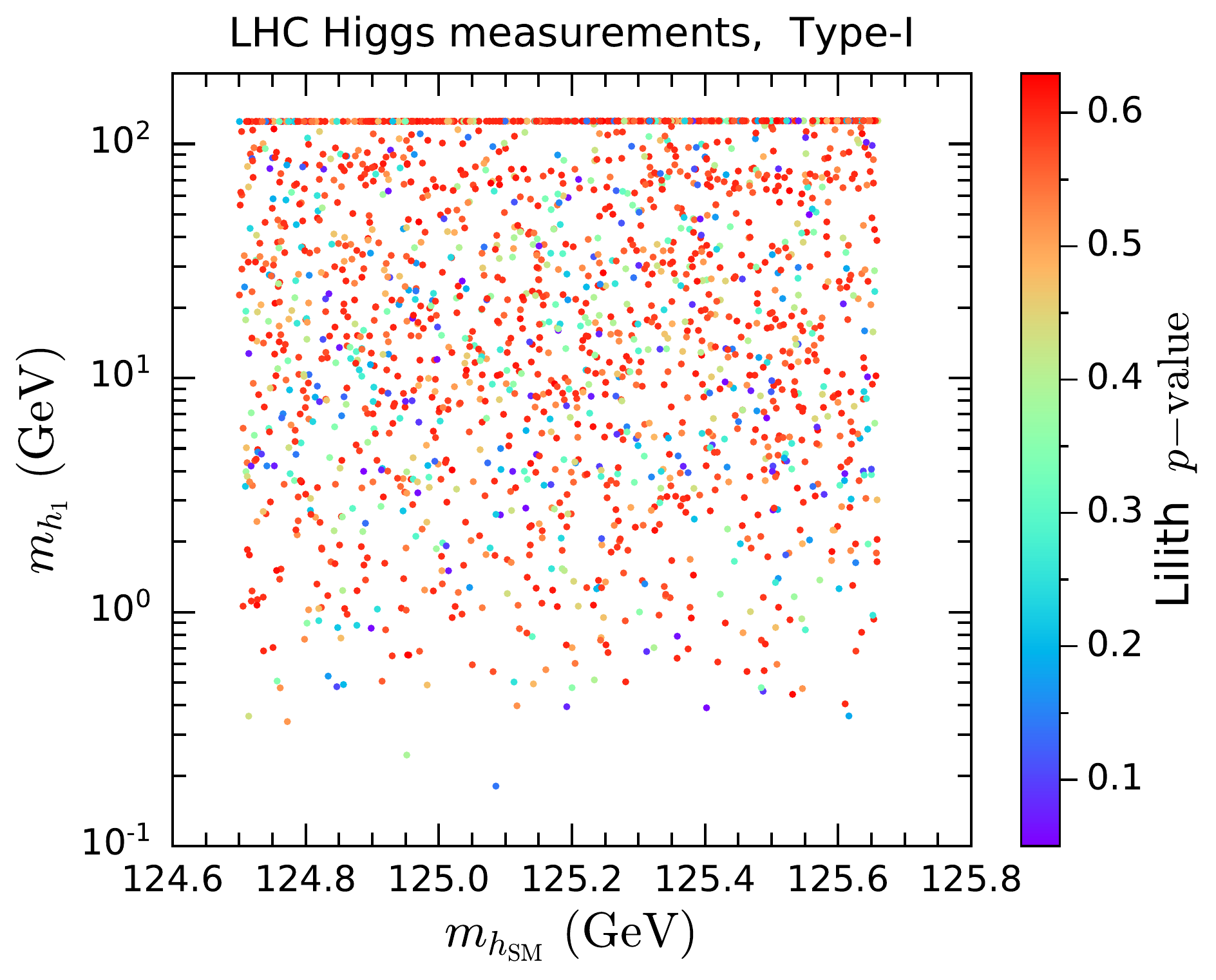}}
\subfigure[~$m_{h_2}$-$m_{h_3}$ plane.\label{fig:mh2-mh3}]
{\includegraphics[width=0.48\textwidth]{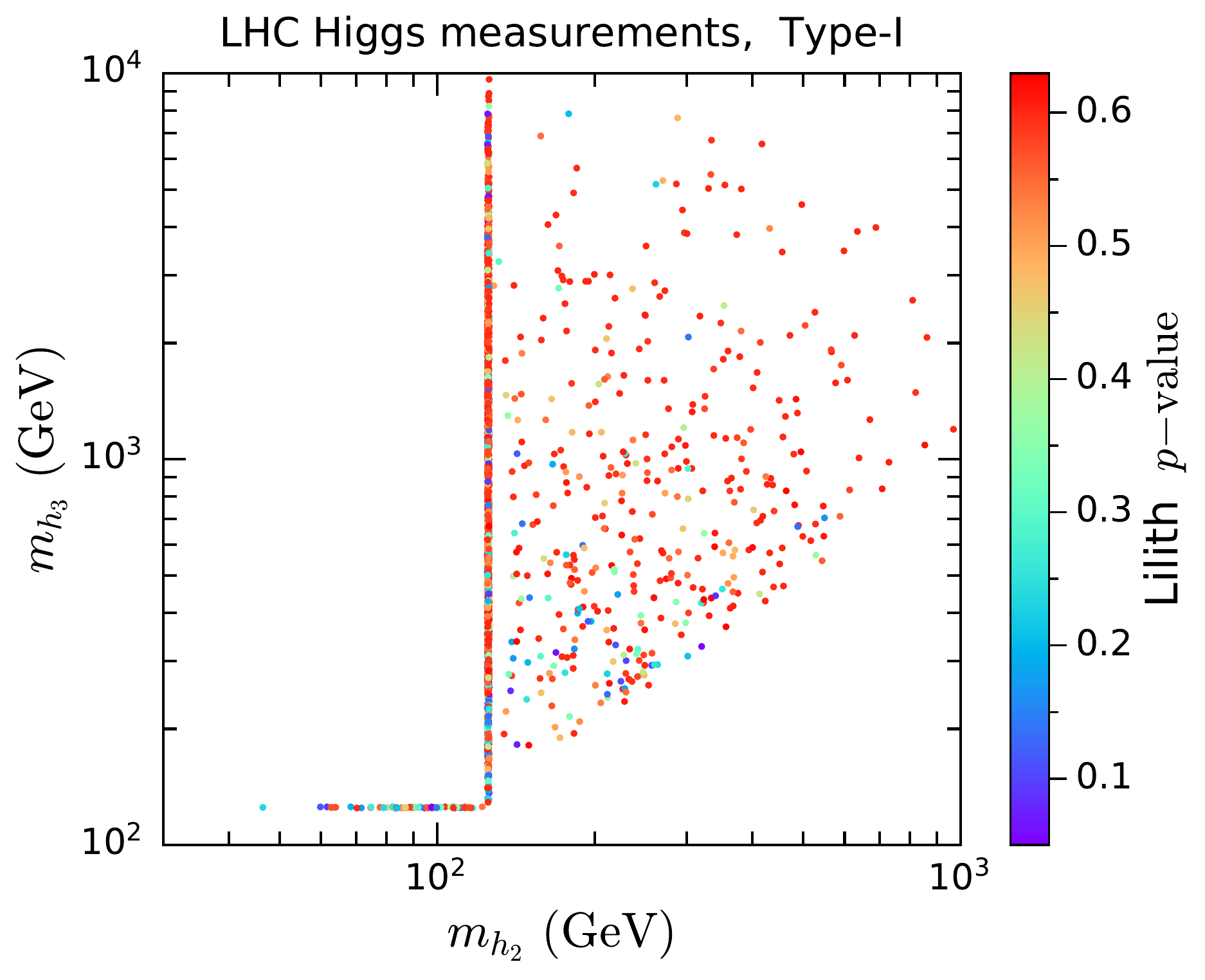}}
\caption{\texttt{Lilith} $p$-values for the selected parameter points projected in the $m_{h_\mathrm{SM}}$-$m_{h_1}$ (a) and $m_{h_2}$-$m_{h_3}$ (b) planes.}
\label{fig:mhSM_mh1_mh2_mh3}
\end{figure}

Additionally, we project the parameter points in the $m_{h_\mathrm{SM}}$-$m_{h_1}$ and $m_{h_2}$-$m_{h_3}$ planes in Figs.~\ref{fig:mhSM-mh1} and \ref{fig:mh2-mh3}, respectively.
In Fig.~\ref{fig:mhSM-mh1}, the points with $h_\mathrm{SM}=h_1$ align along a horizontal line with $m_{h_1} \simeq 125~\si{GeV}$, while the remaining points indicate that the SM-like Higgs boson is not the lightest $CP$-even Higgs boson $h_1$.
On the other hand, two sets of aligned points in Fig.~\ref{fig:mh2-mh3} correspond to $h_\mathrm{SM}=h_2$ and $h_\mathrm{SM}=h_3$.

\begin{figure}[!t]
\centering
\includegraphics[width=0.48\textwidth]{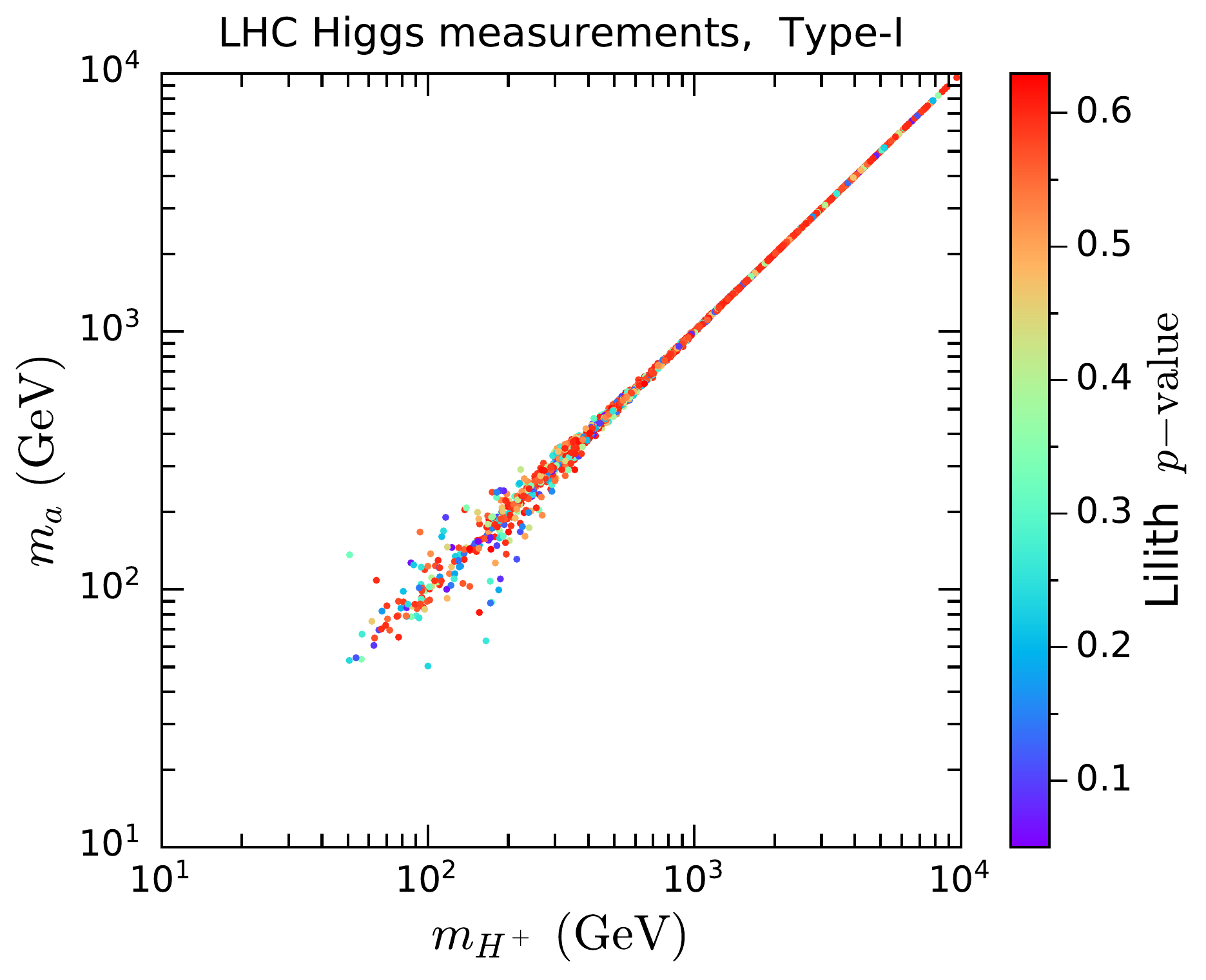}
\caption{\texttt{Lilith} $p$-values for the selected parameter points projected in the $m_{H^+}$-$m_a$ plane.}
\label{fig:mHp-ma}
\end{figure}

The projection on the $m_{H^+}$-$m_a$ plane is presented in Fig.~\ref{fig:mHp-ma}.
From Eq.~\eqref{eq:mHp_ma}, we know that the difference between the masses of the charged Higgs boson $H^+$ and the $CP$-odd Higgs boson $a$ are due to the $\lambda_4$ and $\lambda_5$ couplings.
If ${m}_{12}^2$ is much larger than the $\lambda_4$ and $\lambda_5$ contributions, the difference would be negligible, as demonstrated in Fig.~\ref{fig:mHp-ma} for $m_{H^+}, m_a \gtrsim 500~\si{GeV}$.

\begin{figure}[!t]
\centering
\subfigure[~$|\kappa_V|$-$|\kappa_f|$ plane.\label{fig:kapV-kapf}]
{\includegraphics[width=0.48\textwidth]{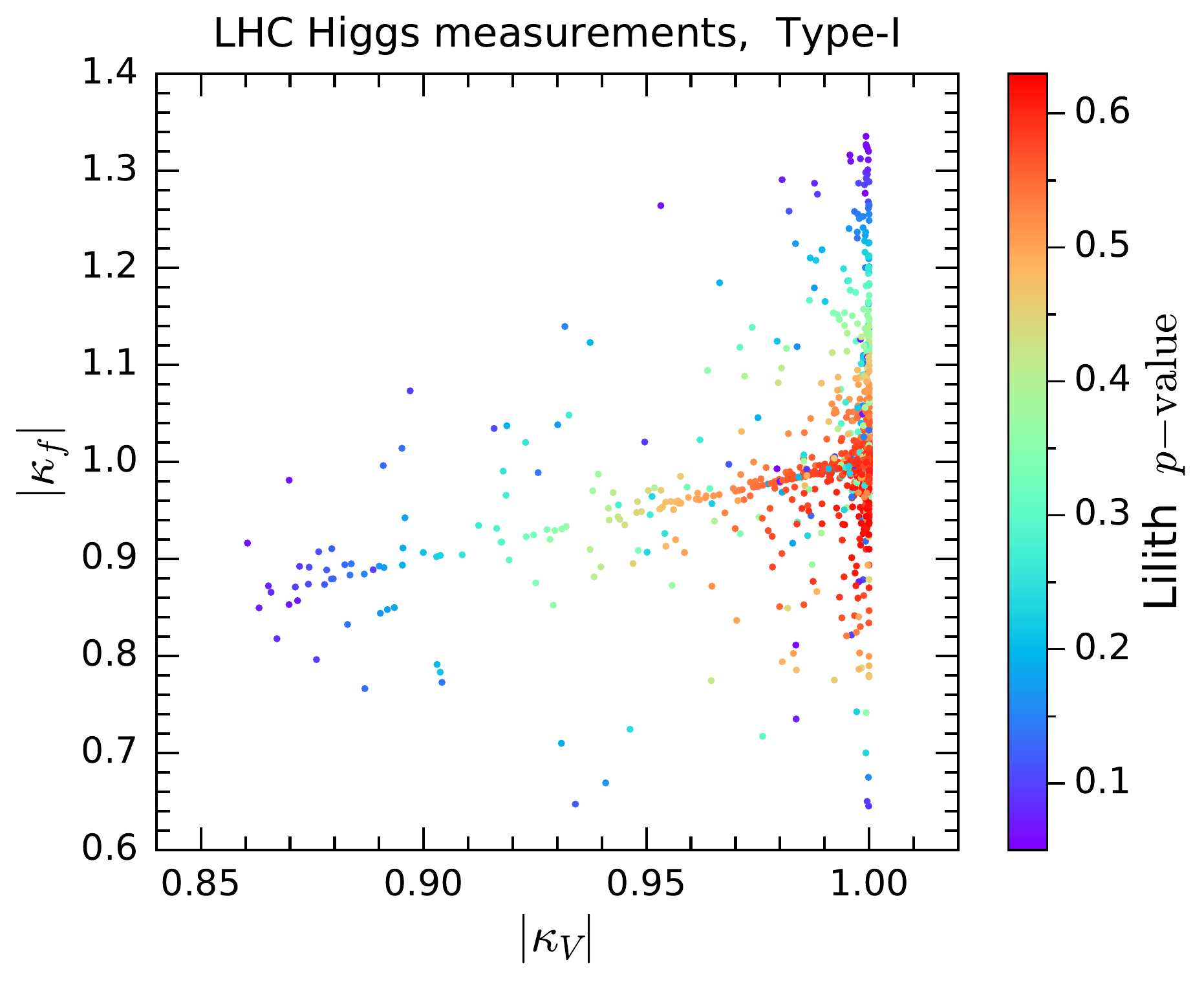}}
\subfigure[~$|O_{1i}|/c_\beta$-$|O_{2i}|/s_\beta$ plane.\label{fig:O1i-O2i}]
{\includegraphics[width=0.48\textwidth]{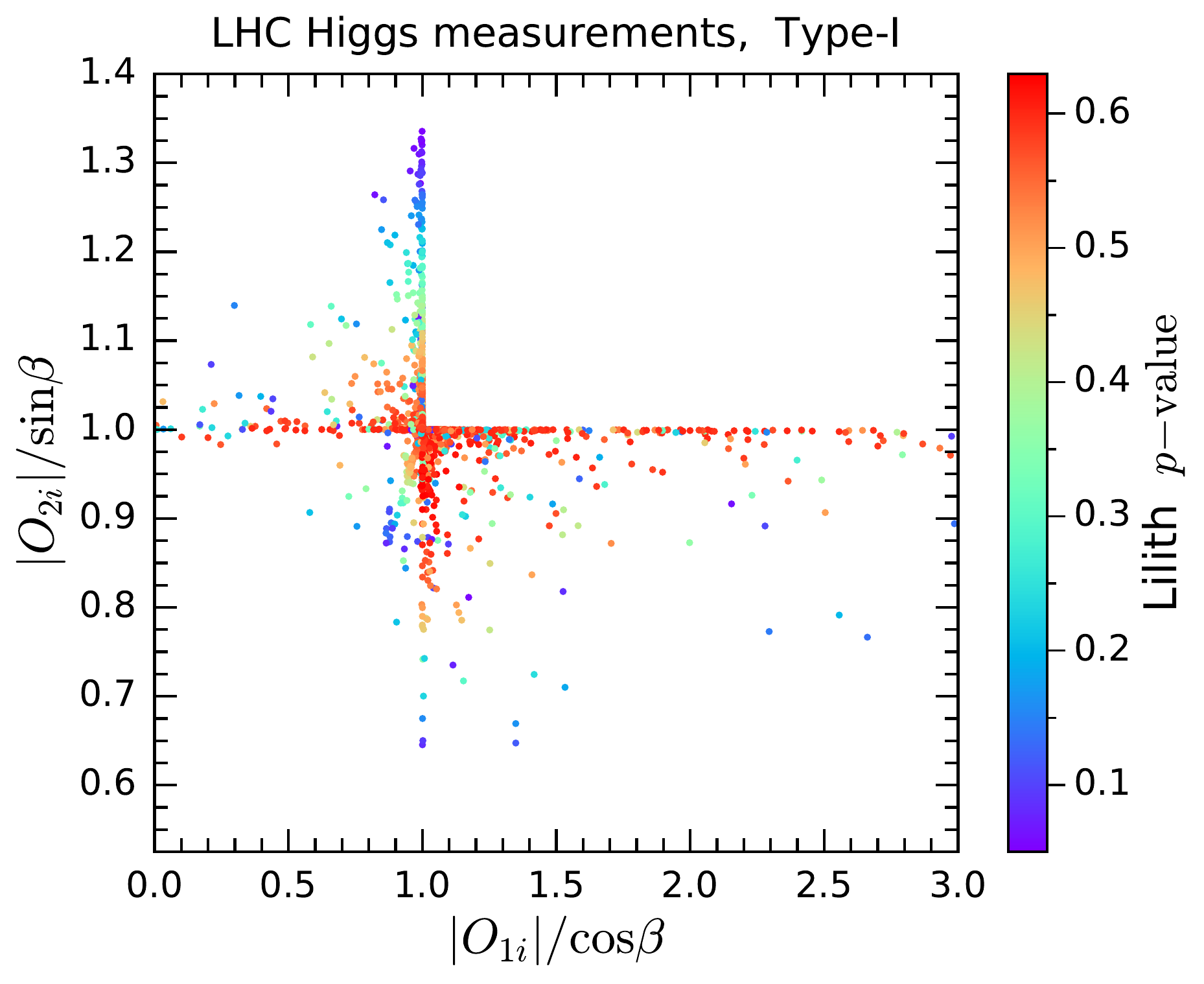}}
\caption{\texttt{Lilith} $p$-values for the selected parameter points projected in the $|\kappa_V|$-$|\kappa_f|$ (a) and $|O_{1i}|/c_\beta$-$|O_{2i}|/s_\beta$ (b) planes.}
\label{fig:kapV_kapf_O1i_O2i}
\end{figure}

Figure~\ref{fig:kapV-kapf} shows the projection on the $|\kappa_V|$-$|\kappa_f|$ plane.
We find that the parameter points with $|\kappa_V| \simeq |\kappa_f| \simeq 1$ have the largest $p$-values, implying that current data still favor that the $125~\si{GeV}$ Higgs boson has SM-like couplings.
Nonetheless, $|\kappa_V|$ may range from $\sim 0.85$ to $\sim 1$, and $|\kappa_f|$ may range from $\sim 0.6$ to $\sim 1.3$.
In addition, there are two categories of parameter points approximately aligning along two outstanding lines.
\begin{itemize}
\item \textbf{Category~1}:
One line in Fig.~\ref{fig:kapV-kapf} corresponds to $|\kappa_V| \simeq |\kappa_f|$.
Actually, the signs of $\kappa_V$ and $\kappa_f$ are the same for all selected parameter points.
This line is thus related to $\kappa_V \simeq \kappa_f$.
The main reason is that if $\tan\beta\gg 1$, we have $s_\beta \simeq 1$ and $c_\beta \simeq 0$, and Eqs.~\eqref{eq:kapV} and \eqref{eq:kapf} become $\kappa_V \simeq \kappa_f \simeq O_{2i}$, where $\kappa_V$ and $\kappa_f$ have a nearly total positive correlation.
As $|O_{2i}| \leq 1$, in this case both $|\kappa_V|$ and $|\kappa_f|$ cannot exceed one.
Most of the parameter points in this category correspond to the horizontal line with $|O_{2i}|/s_\beta \simeq 1$ in the $|O_{1i}|/c_\beta$-$|O_{2i}|/s_\beta$ plane shown in Fig.~\ref{fig:O1i-O2i}, while the rest give $|O_{2i}|/s_\beta < 1$.
\item \textbf{Category~2}:
Another line in Fig.~\ref{fig:kapV-kapf} corresponds to $|\kappa_V| \simeq 1$ with varying $|\kappa_f|$.
This category is related to the vertical line with $|O_{1i}|/c_\beta \simeq 1$ in Fig.~\ref{fig:O1i-O2i}.
From Eq.~\eqref{eq:kapV}, we know that $|O_{1i}| \simeq c_\beta$ and $|O_{2i}| \simeq s_\beta$ could lead to $|\kappa_V| \simeq c_\beta^2 + s_\beta^2 = 1$.
Nonetheless, the second relation $|O_{2i}| \simeq s_\beta$ is not important to keep $|\kappa_V| \simeq 1$ when $s_\beta \ll 1$.
Therefore, in the case of $\tan\beta \ll 1$, $|O_{2i}|/s_\beta$ could deviate from one, resulting in the vertical line in Fig.~\ref{fig:O1i-O2i}.
\end{itemize}
There are some scatter points not belonging in the two categories.
Most of them correspond to $\tan\beta \sim 1$.

\begin{figure}[!t]
\centering
\subfigure[~$\kappa_g$-$\kappa_\gamma$ plane.\label{fig:kapg-kapga}]
{\includegraphics[width=0.48\textwidth]{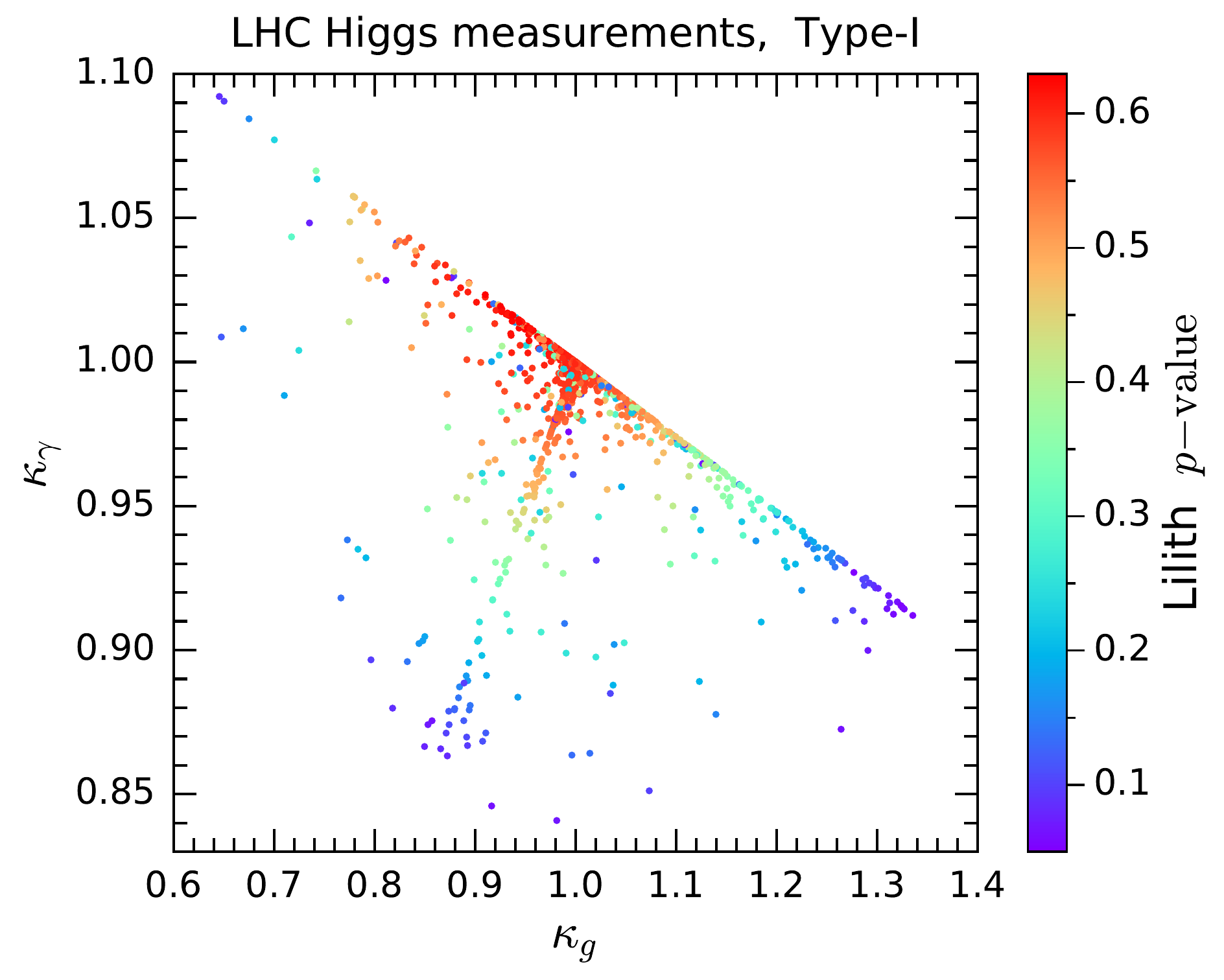}}
\subfigure[~$\kappa_{Z\gamma}$-$\kappa_H$ plane.\label{fig:kapZga-kapH}]
{\includegraphics[width=0.48\textwidth]{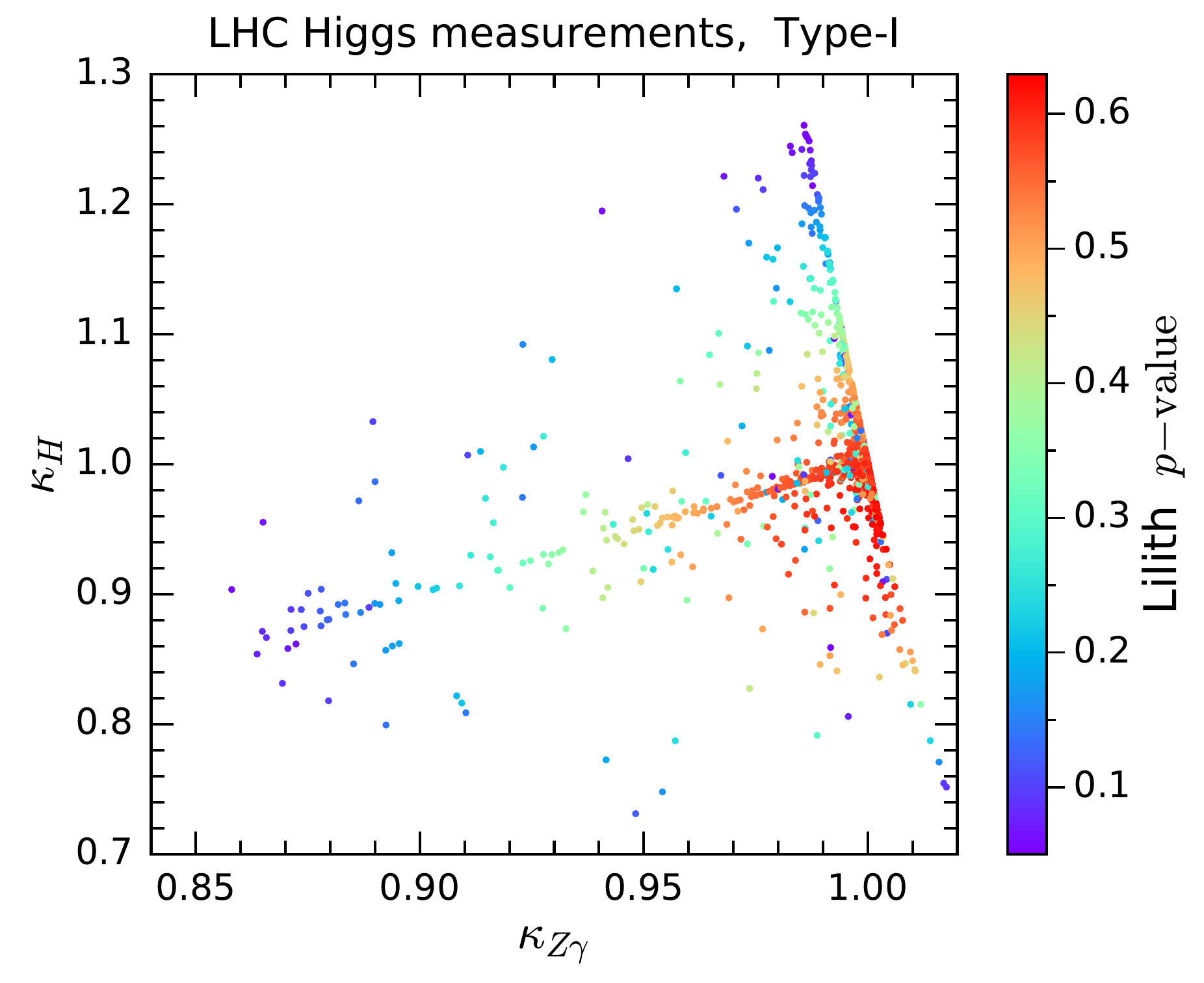}}
\caption{\texttt{Lilith} $p$-values for the selected parameter points projected in the $\kappa_g$-$\kappa_\gamma$ (a) and $\kappa_{Z\gamma}$-$\kappa_H$ (b) planes.}
\label{fig:kap_g_ga_Zga_H}
\end{figure}

The dominant contributions to $\kappa_g$ come from the top and bottom loops, leading to a parametrization of~\cite{Tanabashi:2018oca}
\begin{equation}\label{eq:kap_g}
\kappa _g^2 = 1.06\kappa _t^2 + 0.01\kappa _b^2 - 0.07{\kappa _t}{\kappa _b}.
\end{equation}
On the other hand, $\kappa_\gamma$ is mainly contributed by the $W$ and top loops, resulting in~\cite{Tanabashi:2018oca}
\begin{equation}\label{eq:kap_gamma}
\kappa _\gamma ^2 = 1.59\kappa _W^2 + 0.07\kappa _t^2 - 0.66{\kappa _W}{\kappa _t}.
\end{equation}
In both cases, the interference between the two contributions gives a term with a negative coefficient.
In Fig.~\ref{fig:kapg-kapga}, we project the parameter points in the $\kappa_g$-$\kappa_\gamma$ plane, where the points also align along two lines.
One line implies a positive correlation between $\kappa_g$ and $\kappa_\gamma$, corresponding to category~1.
This is because the relation $\kappa_V \simeq \kappa_f$ gives rise to such a positive correlation via Eqs.~\eqref{eq:kap_g} and \eqref{eq:kap_gamma}.
On the other hand, when $|\kappa_V| \simeq 1$, $\kappa _\gamma$ is negatively correlated to $|\kappa_f|$ as all selected parameter points satisfy $\kappa_V \kappa_f>0$.
As $\kappa_g$ is positively correlated to $|\kappa_f|$, category~2 results in a second line with a negative correlation between $\kappa_g$ and $\kappa_\gamma$.

$\kappa_{Z\gamma}$ is also dominantly contributed by the $W$ and top loops, given by~\cite{Tanabashi:2018oca}
\begin{equation}
\kappa _{Z\gamma }^2 = 1.12\kappa _W^2 + 0.03\kappa _t^2 - 0.15{\kappa _W}{\kappa _t}.
\end{equation}
The correlations of $\kappa_{Z\gamma}$ to $\kappa_V$ and to $\kappa_f$ are similar to those of $\kappa_\gamma$.
In addition, $\kappa _H$ can be expressed as~\cite{Tanabashi:2018oca}
\begin{equation}
\kappa _H^2 = 0.57\kappa _b^2 + 0.06\kappa _\tau ^2 + 0.03\kappa _c^2 + 0.22\kappa _W^2 + 0.03\kappa _Z^2 + 0.09\kappa _g^2 + 0.0023\kappa _\gamma ^2,
\end{equation}
where all the coefficients are positive.
Thus, $\kappa_H$ is positively correlated to both $|\kappa_V|$ and $|\kappa_f|$.
The projection in the $\kappa_{Z\gamma}$-$\kappa_H$ plane are shown in Fig.~\ref{fig:kapZga-kapH}.
Analogous to Fig.~\ref{fig:kapg-kapga}, category~1 leads to
a line indicating a positive correlation between $\kappa_{Z\gamma}$ and $\kappa_H$ in Fig.~\ref{fig:kapZga-kapH}.
Besides, parameter points in category~2 roughly align along a second line with a negative correlation.

\begin{figure}[!t]
\centering
\subfigure[~$m_\chi$-$\mathrm{BR}_\mathrm{inv}$ plane.\label{fig:mchi-BRinv}]
{\includegraphics[width=0.48\textwidth]{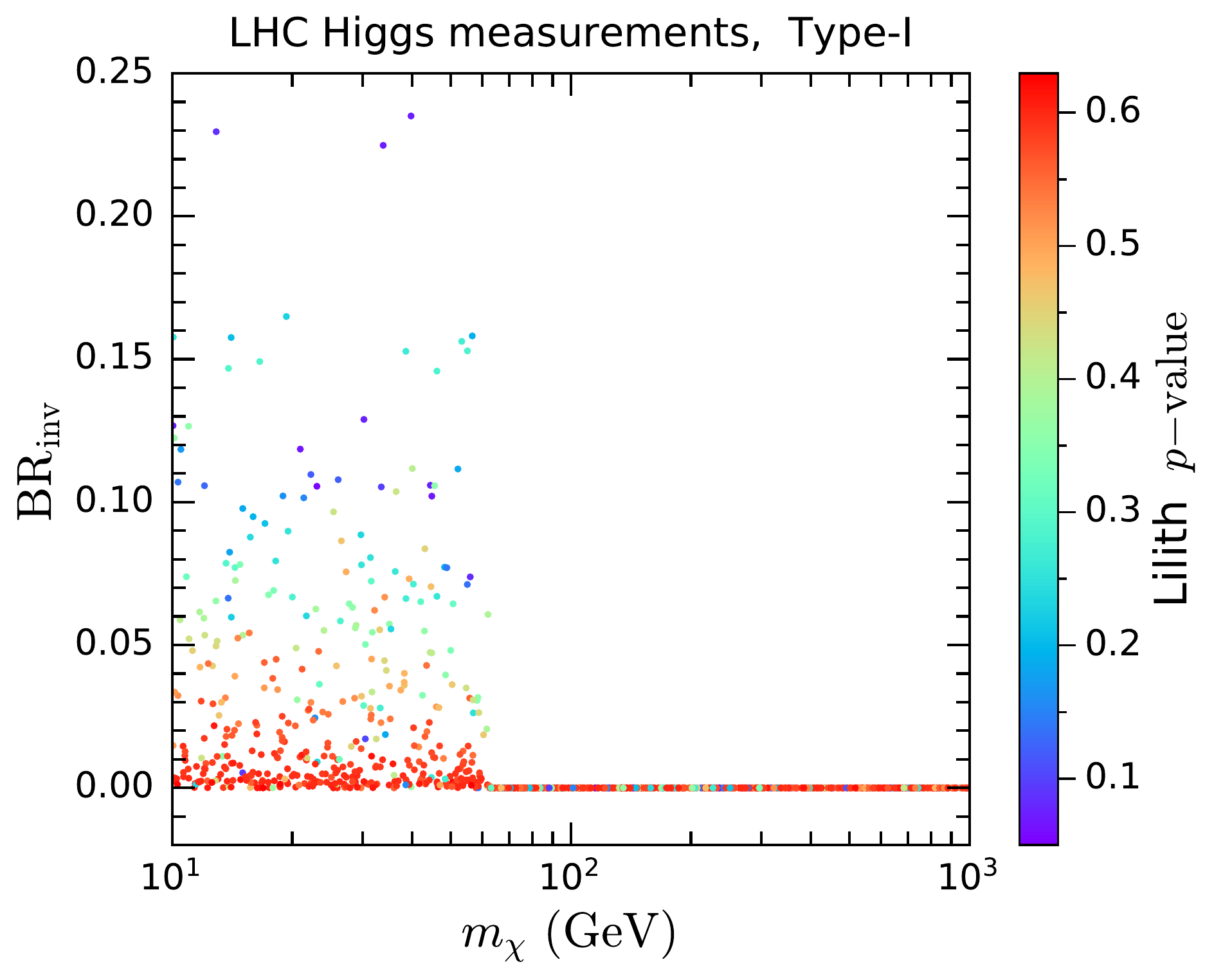}}
\subfigure[~$\Gamma_{h_\mathrm{SM}}$-$\mathrm{BR}_\mathrm{und}$ plane.\label{fig:width-BRund}]
{\includegraphics[width=0.48\textwidth]{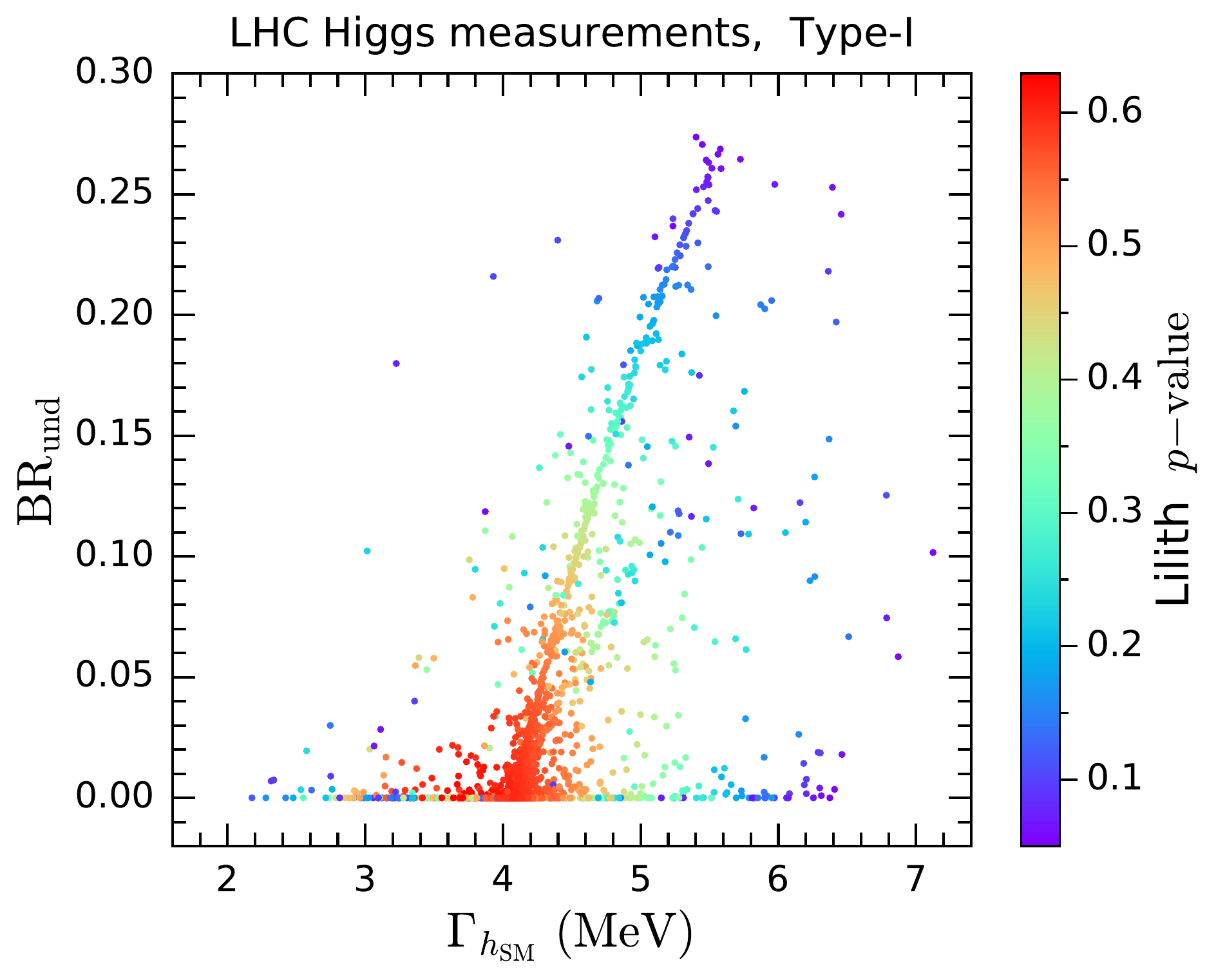}}
\caption{\texttt{Lilith} $p$-values for the selected parameter points projected in the $m_\chi$-$\mathrm{BR}_\mathrm{inv}$ (a) and $\Gamma_{h_\mathrm{SM}}$-$\mathrm{BR}_\mathrm{und}$ (b) planes.}
\label{fig:mchi_BRinv_width_BRund}
\end{figure}

In Fig.~\ref{fig:mchi-BRinv}, we show the projection in the $m_\chi$-$\mathrm{BR}_\mathrm{inv}$ plane.
When $m_\chi > m_{h_\mathrm{SM}}/2$, we have $\mathrm{BR}_\mathrm{inv}=0$, because the invisible decay $h_\mathrm{SM}\to \chi\chi$ is kinematically forbidden.
When $m_\chi < m_{h_\mathrm{SM}}/2$, the invisible branching ratio $\mathrm{BR}_\mathrm{inv}$ could be as large as $\sim 25\%$ and still consistent with data at 95\% C.L.
The projection in the $\Gamma_{h_\mathrm{SM}}$-$\mathrm{BR}_\mathrm{und}$ plane is presented in Fig.~\ref{fig:width-BRund}.
We find that the undetected BSM branching ratio $\mathrm{BR}_\mathrm{und}$ can be allowed up to $\sim 27\%$, while the total width $\Gamma_{h_\mathrm{SM}}$ can range from $\sim 2$ to $\sim 7~\si{MeV}$.
There is a line implying a positive correlation between $\Gamma_{h_\mathrm{SM}}$ and $\mathrm{BR}_\mathrm{und}$.
This is reasonable, because opening new decay channels enlarges the total width.

\begin{figure}[!t]
\centering
\subfigure[~$\tan\beta$-$\tilde{\lambda}_6$ plane.\label{fig:tanb-lam6t}]
{\includegraphics[width=0.48\textwidth]{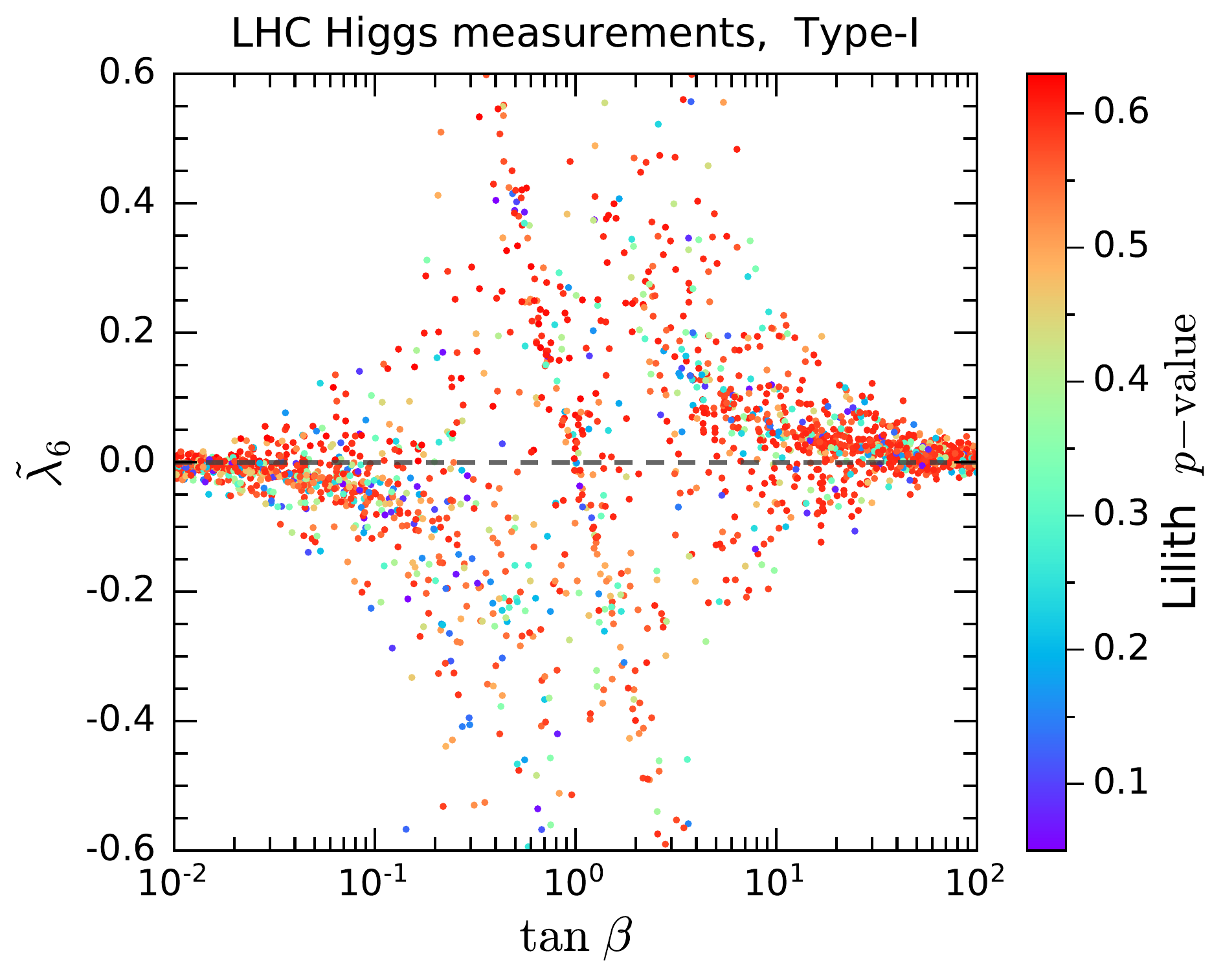}}
\subfigure[~$\tan\beta$-$\tilde{\kappa}_1$ plane.\label{fig:tanb-kap1t}]
{\includegraphics[width=0.48\textwidth]{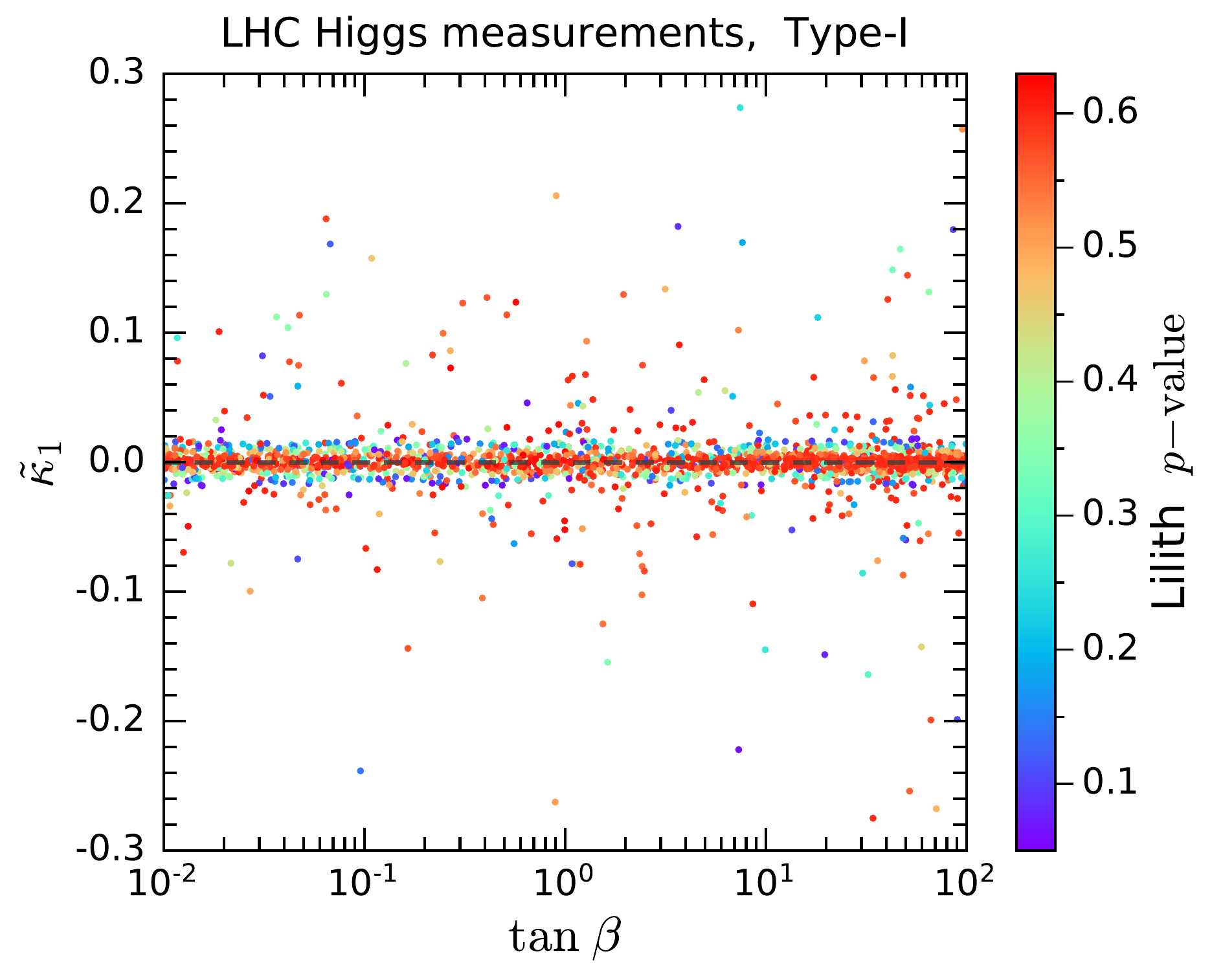}}
\caption{\texttt{Lilith} $p$-values for the selected parameter points projected in the $\tan\beta$-$\tilde{\lambda}_6$ (a) and $\tan\beta$-$\tilde{\kappa}_1$ (b) planes.
The dashed lines indicate the alignment limit.}
\label{fig:tanb-lam6t_kap1t}
\end{figure}

In order to investigate the alignment limit, which corresponds to $\tilde{\lambda}_6 = \tilde{\kappa}_1 = 0$, the selected parameter points are projected in the $\tan\beta$-$\tilde{\lambda}_6$ and $\tan\beta$-$\tilde{\kappa}_1$ planes in Figs.~\ref{fig:tanb-lam6t} and \ref{fig:tanb-kap1t}, respectively.
We find that most of the selected points satisfy $\tilde{\kappa}_1 \simeq 0$, showing no particular dependence on $\tan\beta$.
On the other hand, $\tilde{\lambda}_6$ is typically close to zero for $\tan\beta \gtrsim 20$ and $\tan\beta \lesssim 0.05$.
For $0.05 \lesssim \tan\beta \lesssim 20$, there is no particular favor in the alignment limit.

\subsection{DM relic abundance}

The thermal relic abundance of dark matter is essentially determined by the total velocity-averaged annihilation cross section at the freeze-out epoch, which we denote as $\svann_\mathrm{FO}$.
In our model, the DM candidate $\chi$ has the following annihilation channels if kinematically allowed.
\begin{itemize}
\item Annihilation into a pair of fermions, $\chi\chi\to f\bar{f}$.
This channel is mediated by $s$-channel $CP$-even Higgs bosons and suppressed by fermion masses.
Thus, $t\bar{t}$ and $b\bar{b}$ are the important final states.
\item Annihilation into a pair of weak gauge bosons, $\chi\chi\to W^+ W^-, ZZ$.
This channel is also mediated by $s$-channel $CP$-even Higgs bosons.
\item Annihilation into a weak gauge boson and a Higgs boson, $\chi\chi\to W^\pm H^\mp, Za$, mediated by $s$-channel $CP$-even Higgs bosons.
\item Annihilation into a pair of $CP$-even Higgs bosons, $\chi\chi\to h_i h_j$.
This channel can be mediated by $s$-channel $CP$-even Higgs bosons, as well as by $t$- and $u$-channel $\chi$.
Additionally, there are contributions from quartic scalar couplings.
\item Annihilation into a pair of $CP$-odd or charged Higgs bosons, $\chi\chi\to aa, H^+H^-$.
This channel is contributed by the mediation of $s$-channel $CP$-even Higgs bosons and quartic scalar couplings.
\end{itemize}

Some numerical tools are adopted to calculate the relic abundance.
We implement the model with \texttt{FeynRules~2.3.34}~\cite{Alloul:2013bka}, and import the generated model files to a Monte Carlo generator \texttt{MadGraph5\_aMC@NLO~2.6.5}~\cite{Alwall:2014hca}.
Then we utilize a \texttt{MadGraph} plugin \texttt{MadDM~3}~\cite{Ambrogi:2018jqj} to compute the relic abundance $\Omega_\chi h^2$ for each parameter point.

\begin{figure}[!t]
\centering
\includegraphics[width=0.48\textwidth]{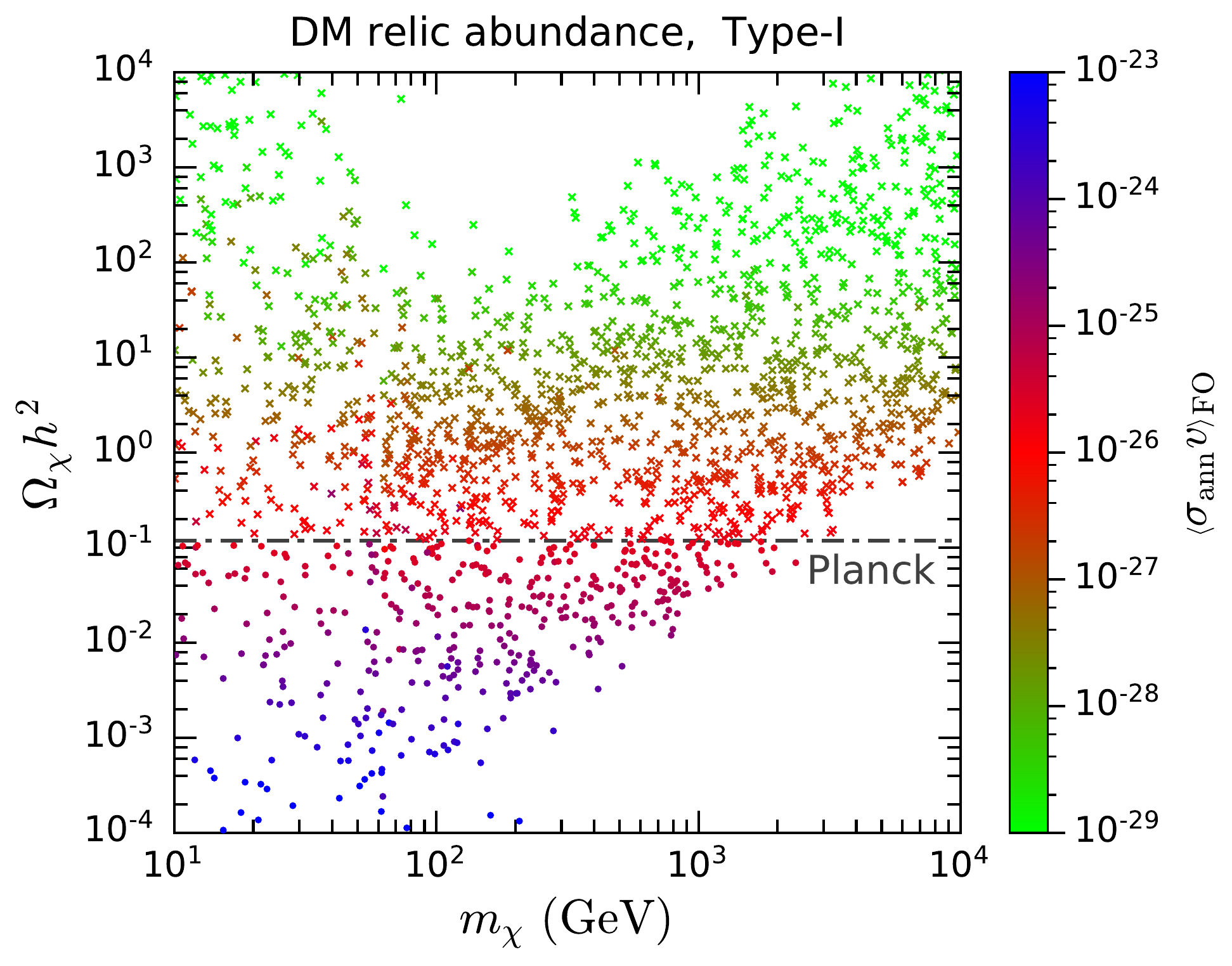}
\caption{Selected parameter points projected in the $m_\chi$-$\Omega_\chi h^2$ plane.
The colors indicate the freeze-out annihilation cross section $\svann_\mathrm{FO}$.
The dot-dashed line corresponds to the relic abundance measured by Planck~\cite{Ade:2015xua}.
The parameter points denoted with crosses lead to DM overproduction, while those denoted with dots are allowed by data.}
\label{fig:mchi-Omegah2}
\end{figure}

The relic abundance predicted by the selected parameter points is shown in Fig.~\ref{fig:mchi-Omegah2}, where the color bar denotes the freeze-out annihilation cross section $\svann_\mathrm{FO}$.
We find that the observed value $\Omega h^2 = 0.1186\pm 0.0020$ given by the Planck experiment~\cite{Ade:2015xua} corresponds to $\svann_\mathrm{FO}\sim \mathcal{O}(10^{-26})~\si{cm^3/s}$, which is typical for thermal dark matter.
Increase in $m_\chi$ typically reduces the annihilation cross section and hence increases the relic abundance.
Consequently, if the DM candidate is too heavy, say $m_\chi \gtrsim 3~\si{TeV}$, the observed relic abundance could not be achieved.

In Fig.~\ref{fig:mchi-Omegah2}, the parameter points predicting $\Omega_\chi h^2$ over the observed value by $2\sigma$ are denoted with crosses.
These points are considered to be excluded by data, because DM overproduction by the thermal mechanism contradicts standard cosmology.
On the other hand, if the predicted thermal relic abundance is too low, there could be some nonthermal production~\cite{Lin:2000qq,Fujii:2002kr} occurring after DM freezes out.

\subsection{Indirect detection}

In this subsection, we discuss constraints from $\gamma$-ray indirect detection experiments.
There are couples of dwarf spheroidal galaxies discovered as satellites of the Milky Way Galaxy.
They are considered as the largest substructures of the Galactic dark halo, predicted by the cold DM scenario~\cite{Springel:2008cc,Diemand:2008in}.
As known so far, they are the most DM-dominated systems~\cite{Strigari:2013iaa}.
Moreover, $\gamma$-ray emissions from typical astrophysical sources, such as neutral and ionized gases and recent star formation activity, are expected to be rare in such dwarf galaxies~\cite{Mateo:1998wg,Gallagher:2003nx,Grcevich:2009gt}.
These properties make them perfect targets for searching for $\gamma$-ray emissions from DM annihilation.

The DM velocity dispersion in dwarf galaxies is typically $\sim \mathcal{O}(10^{-5})$~\cite{Walker:2009zp}, which is smaller than DM velocities at the freeze-out epoch by 4 orders of magnitude.
Therefore, if the velocity dependence is significant in DM annihilation, the total velocity-averaged cross section in dwarf galaxies $\svann_\mathrm{dwarf}$ could be much different from the freeze-out value $\svann_\mathrm{FO}$.

\begin{figure}[!t]
\centering
\subfigure[~$m_\chi$-$\svann_\mathrm{dwarf}/\svann_\mathrm{FO}$ plane.\label{fig:mchi-svratio}]
{\includegraphics[width=0.48\textwidth]{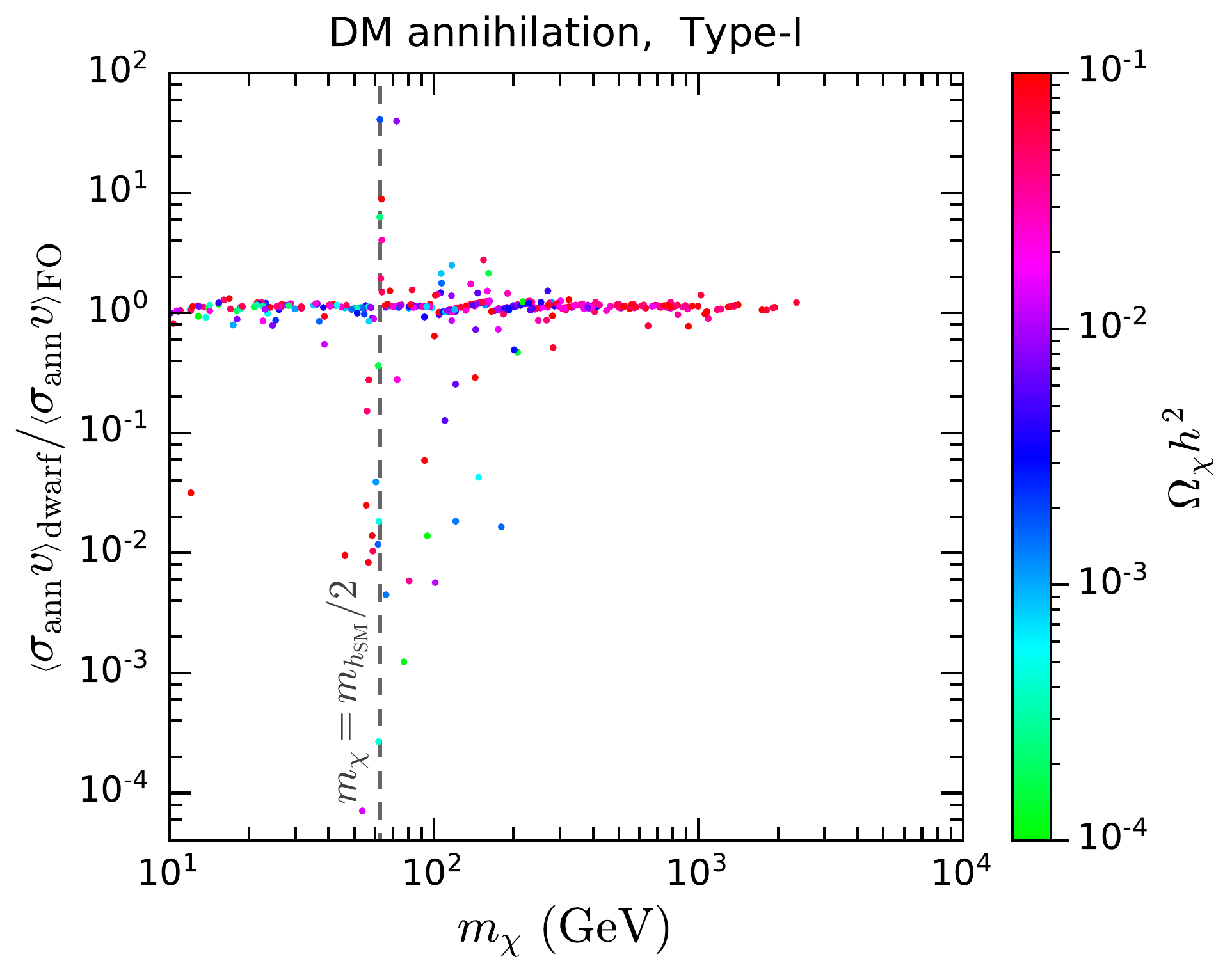}}
\subfigure[~$m_\chi$-$\svann_\mathrm{dwarf}$ plane.\label{fig:mchi-svdwarf}]
{\includegraphics[width=0.48\textwidth]{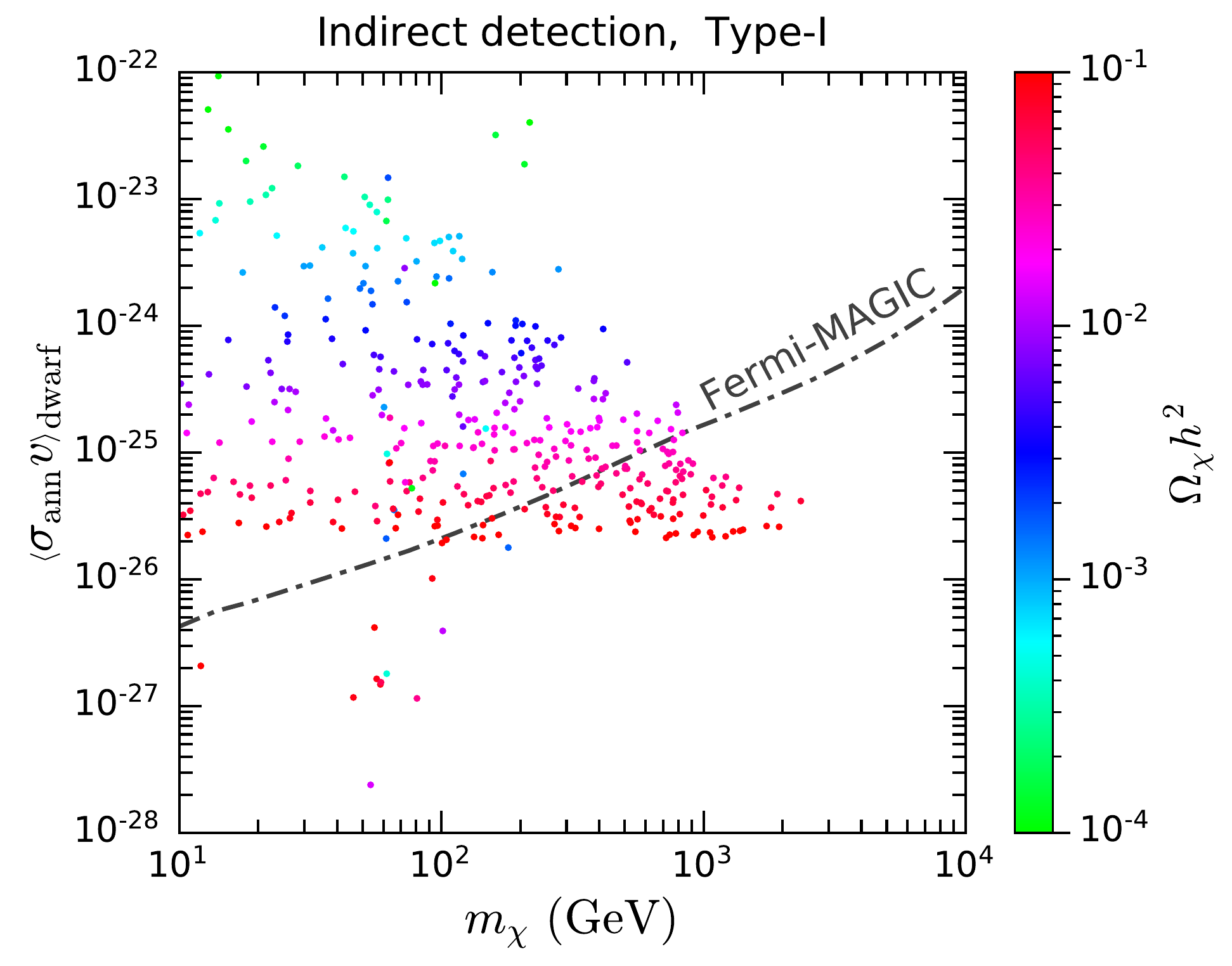}}
\caption{Parameter points with data-allowed relic abundance projected in the $m_\chi$-$\svann_\mathrm{dwarf}/\svann_\mathrm{FO}$ (a) and $m_\chi$-$\svann_\mathrm{dwarf}$ (b) planes.
The colors indicate the predicted relic abundance $\Omega_\chi h^2$.
The dashed line in the left panel denotes the location of $m_\chi = m_{h_\mathrm{SM}}/2$.
The dot-dashed line in the right panel denotes the 95\% C.L. upper limits from $\gamma$-ray observations of dwarf galaxies by Fermi-LAT and MAGIC~\cite{Ahnen:2016qkx}.}
\label{fig:mchi_sv}
\end{figure}

We further use \texttt{MadDM} to calculate $\svann_\mathrm{dwarf}$ for each parameter point assuming the average DM velocity is $2\times 10^{-5}$.
The ratio of $\svann_\mathrm{dwarf}$ to $\svann_\mathrm{FO}$ is demonstrated in Fig.~\ref{fig:mchi-svratio}, where the parameter points excluded by the Planck relic abundance measurement are not shown.
Most of the parameter points give $\svann_\mathrm{dwarf}/\svann_\mathrm{FO} \sim 1$, indicating that $s$-wave annihilation is dominant.
Nonetheless, some points give the ratio away from $\mathcal{O}(1)$, indicating significant dependence on velocity.
This is typically due to DM annihilation through the resonances of $CP$-even Higgs bosons,
since the resonance effect extremely depends on the difference between the resonance location and the velocity-dependent center-of-mass energy~\cite{Gondolo:1990dk,Griest:1990kh}.

The vertical dashed line in Fig.~\ref{fig:mchi-svratio} indicates the location of $m_\chi = m_{h_\mathrm{SM}}/2$, corresponding to the resonance of the SM-like Higgs boson.
We can see that the ratio around this line could range from $\sim 10^{-4}$ to $\sim 30$.
On the other hand, locations of the other resonances are not fixed, but their effects are also important.

Figure~\ref{fig:mchi-svdwarf} shows the projection of the parameter points in the $m_\chi$-$\svann_\mathrm{dwarf}$ plane, as well as the 95\% C.L. upper limits on $\svann_\mathrm{dwarf}$ given by an analysis of Fermi-LAT and MAGIC $\gamma$-ray observations~\cite{Ahnen:2016qkx}.
The analysis combined 6-yr observations of 15 dwarf galaxies from the Fermi-LAT satellite experiment and 158-hr observations of a single dwarf galaxy Segue~1 from the MAGIC Cherenkov telescopes, assuming that the DM distributions in the dwarf galaxies follow the Navarro-Frenk-White profile~\cite{Navarro:1996gj}.
The limits were obtained assuming that DM solely annihilates into $b\bar{b}$.
However, there are various DM annihilation channels in our model.
Fortunately, the $\gamma$-ray spectra yielded from these channels should be similar to the spectrum from the $b\bar{b}$ channel, because they are contributed by similar processes, such as hadronization, hadron decays, and final state radiation.
Therefore, we have a good reason to expect that the $b\bar{b}$ limits are approximately applicable to our case.

Some parameter points shown in Fig.~\ref{fig:mchi-svdwarf} predict a thermal relic abundance lower than the Planck observed value.
In this case, the interpretation of indirect detection constraints depends on the assumption of DM composition.
If we assume the DM candidate $\chi$ in our model makes up all dark matter in the Universe, nonthermal production~\cite{Lin:2000qq,Fujii:2002kr} would be needed to realize the observed abundance.
Under such an assumption, the Fermi-MAGIC constraint on the parameter space can be directly read off from Fig.~\ref{fig:mchi-svdwarf}.
We can observe that a large fraction of the parameter points with $m_\chi\lesssim 1~\si{TeV}$ are ruled out, while the parameter points with $m_\chi\gtrsim 100~\si{GeV}$ and $\Omega_\chi h^2 \sim 0.1$ are not excluded.
Additionally, if $m_\chi \simeq m_{h_\mathrm{SM}}/2$, the resonance effect could both yield a data-allowed relic abundance and lead to a small $\svann_\mathrm{dwarf}$ evading the indirect detection constraint.

\begin{figure}[!t]
\centering
\includegraphics[width=0.48\textwidth]{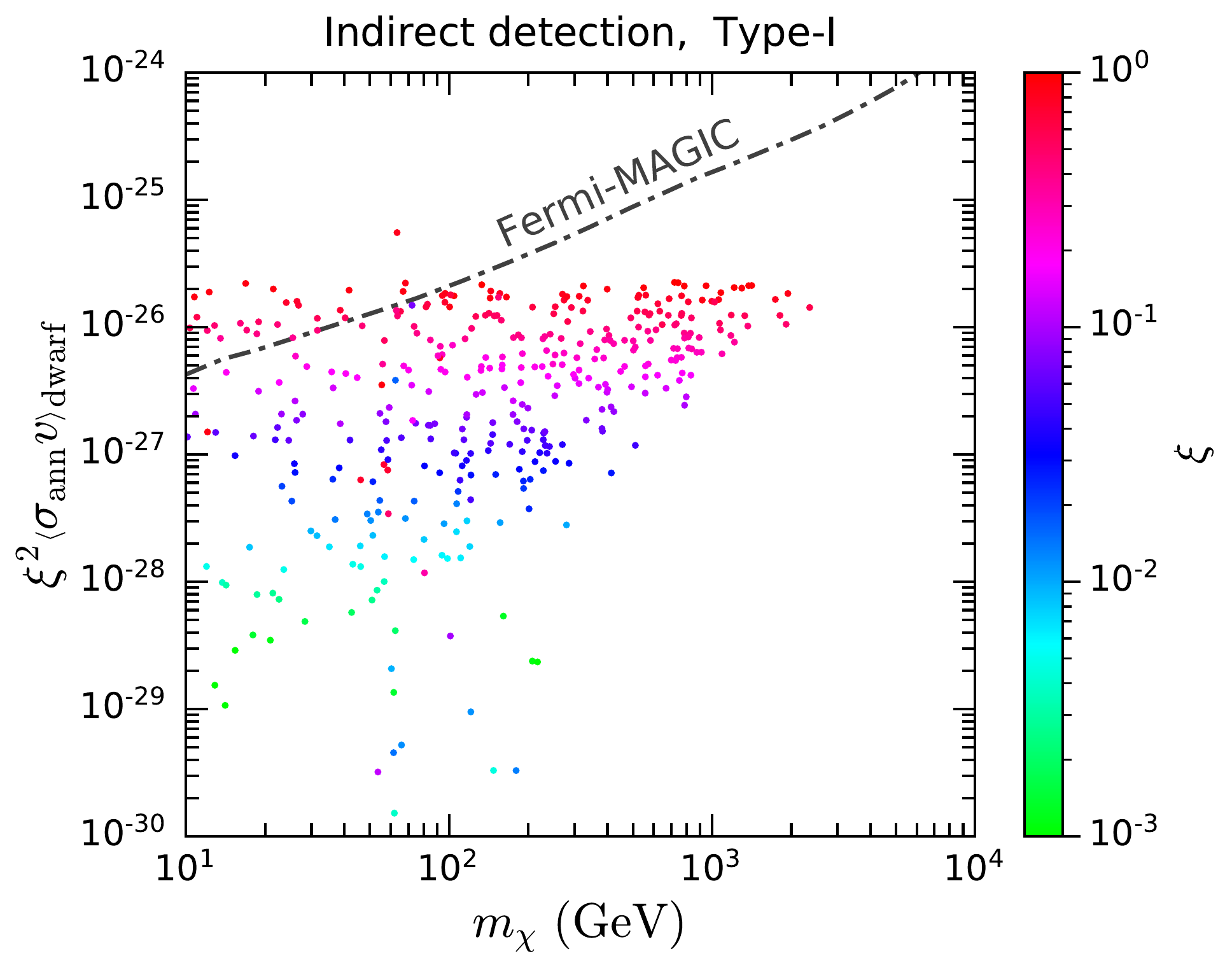}
\caption{Parameter points projected in the $m_\chi$-$\xi^2\svann_\mathrm{dwarf}$ plane.
The colors indicate the ratio of the predicted relic abundance to the observed value, $\xi$.
The dot-dashed line denotes the 95\% C.L. upper limits from Fermi-MAGIC $\gamma$-ray observations of dwarf galaxies~\cite{Ahnen:2016qkx}.}
\label{fig:mchi_xi2sv}
\end{figure}

Another reasonable assumption is that the relic abundance of the DM candidate $\chi$ is exactly predicted by the thermal mechanism, and hence it could only constitute a fraction of all dark matter.
The fraction is given by the ratio of the predicted value to the Planck observed value, $\xi\equiv\Omega_\chi/\Omega$.
Thus, the $\chi\chi$ annihilation cross section in dwarf galaxies should be effectively rescaled to $\xi^2 \svann_\mathrm{dwarf}$ for comparing with indirect detection constraints.
Figure~\ref{fig:mchi_xi2sv} presents the parameter points projected in the $m_\chi$-$\xi^2\svann_\mathrm{dwarf}$ plane.
Under this assumption, most of the parameter points evade the Fermi-MAGIC constraint.

\section{Conclusions and outlook}
\label{sec:concl}

In this paper, we have studied the pNGB DM framework with two $\SUtwoL$ Higgs doublets $\Phi_1$ and $\Phi_2$.
The DM candidate $\chi$ is the imaginary part of a complex scalar $S$, which is a SM gauge singlet.
Most of the scalar potential terms respect a global $\Uone$ symmetry $S\to e^{i\alpha} S$, except for a soft breaking term giving mass to $\chi$.
As a result, $\chi$ becomes a stable massive pNGB.
Mass eigenstates in the scalar sector also include three $CP$-even Higgs boson $h_i$, a $CP$-odd Higgs boson $a$, and charged Higgs bosons $H^\pm$.

There are four possible types of Yukawa couplings without tree-level FCNCs, just as in usual two-Higgs-doublet models.
DM scattering off nucleons is mediated by the $CP$-even Higgs bosons.
Because of the pNGB nature of $\chi$, the scattering amplitude vanishes in the limit of zero momentum transfer for all the four Yukawa coupling types.
Although loop corrections lead to a small nonvanishing amplitude, current and near future direct detection experiments are incapable of probing such a DM candidate.

Taking the type-I Yukawa couplings as an example, we have performed a random scan in the 12-dimensional parameter space.
The selected parameter points are required to provide a SM-like Higgs boson whose properties are consistent with current LHC Higgs measurements.
For $\tan\beta \gg 1$ or $\tan\beta \ll 1$, one of the Higgs doublets acts as the SM-like Higgs doublet, i.e., $\Phi_2\simeq \Phi_h$ or $\Phi_1\simeq \Phi_h$, and
most of the selected parameter points satisfy $\tilde{\lambda}_6 \simeq 0$ and $\tilde{\kappa}_1 \simeq 0$, corresponding to the alignment limit.
On the other hand, for $\tan\beta \sim 1$ there is no preference to the alignment limit.

We have also calculated the relic abundance and annihilation cross sections predicted by the selected parameter points.
For $m_\chi \lesssim 3~\si{TeV}$, it is possible to achieve the observed relic abundance.
Because of the resonance effect, the present velocity-averaged annihilation cross section at dwarf galaxies could be rather different from that in the freeze-out epoch.
If we assume that $\chi$ makes up all dark matter in the Universe via thermal and nonthermal mechanisms, Fermi-LAT and MAGIC observations of dwarf galaxies have excluded a large fraction of parameter points with $m_\chi \lesssim 1~\si{TeV}$.
Nonetheless, for $m_\chi \simeq m_{h_\mathrm{SM}}/2$ or $100~\si{GeV} \lesssim m_\chi \lesssim 3~\si{TeV}$, it is still possible to simultaneously satisfy the constraints from the relic abundance observation and indirect detection.
If we assume that $\chi$ only constitutes a fraction of all dark matter when the predicted thermal relic abundance is lower than the observed value, most of the parameter points can evade the Fermi-MAGIC constraint.

Differences among the four types of Yukawa couplings are encoded in the coefficients $\xi_{h_i}^f$ and $\xi_a^f$ given by Table~\ref{tab:xi}, which lead to different expressions for $\kappa_b$ and $\kappa_\tau$.
Both $\kappa_b$ and $\kappa_\tau$ in type II are  different from those in type I.
Thus, we expect the parameter points in type II favored by Higgs measurements should have distinct behaviors from the result present above.
On the other hand, $\kappa_b$ in the lepton-specific (flipped) Yukawa couplings is identical to that in type I (type II), but $\kappa_\tau$ is different.
This could cause minor differences in global fits.

Such a pNGB DM model is strongly related to Higgs physics.
The proposed future Higgs factories, such as CEPC~\cite{CEPCStudyGroup:2018ghi}, ILC~\cite{Baer:2013cma}, and FCC-ee~\cite{Abada:2019zxq}, would greatly improve the Higgs measurements.
We expect that these measurements could significantly restrict the parameter space in our model.
Nevertheless, Higgs measurements are not able to pin down the DM candidate mass $m_\chi$, which is solely determined by the soft breaking term that does not affect the rest of the scalar masses.
Thus, indirect detection experiments in the future are essentially important for exploring this model.

\begin{acknowledgments}

This work is supported in part by the National Natural Science Foundation of China under Grants No.~11805288, No.~11875327, and No.~11905300, the China Postdoctoral Science Foundation
under Grant No.~2018M643282, the Natural Science Foundation of Guangdong Province
under Grant No.~2016A030313313,
the Fundamental Research Funds for the Central Universities,
and the Sun Yat-Sen University Science Foundation.

\end{acknowledgments}

\appendix

\section{Scalar and gauge trilinear couplings}
\label{app:tri_coup}

From the scalar potential \eqref{eq:V}, we derive the scalar trilinear couplings as
\begin{equation}
{\mathcal{L}_{{\mathrm{tri}}}} = \sum\limits_{i = 1}^3 {\left( {\frac{1}{2}{g_{{h_i}{\chi ^2}}}\,{h_i}{\chi ^2} + \frac{1}{2}{g_{{h_i}{a^2}}}\,{h_i}{a^2} + {g_{{h_i}{H^ - }{H^ + }}}\,{h_i}{H^ - }{H^ + }} \right)}  + \sum\limits_{i,j,k = 1}^3 {{g_{ijk}}\,{h_i}{h_j}{h_k}},
\end{equation}
where $g_{{h_i}{\chi ^2}}$ is already given by Eq.~\eqref{eq:g_hchichi}, and the other coupling coefficients are given by
\begin{eqnarray}
{g_{{h_i}{a^2}}} &=&  - \{ [s_\beta ^2{\lambda _1} + c_\beta ^2({\lambda _3} + {\lambda _4} - {\lambda _5})]{v_1} - 2{s_\beta }{c_\beta }{\lambda _5}{v_2}\} {O_{1i}} 
\nonumber\\
&& - \{ [c_\beta ^2{\lambda _2} + s_\beta ^2({\lambda _3} + {\lambda _4} - {\lambda _5})]{v_2} - 2{s_\beta }{c_\beta }{\lambda _5}{v_1}\} {O_{2i}}
 - (s_\beta ^2{\kappa _1} + c_\beta ^2{\kappa _2} ){v_s}{O_{3i}},
\\
{g_{{h_i}{H^ - }{H^ + }}} &=&  - [(s_\beta ^2{\lambda _1} + c_\beta ^2{\lambda _3}){v_1} - {s_\beta }{c_\beta }({\lambda _4} + {\lambda _5}){v_2}]{O_{1i}} 
\nonumber\\
&& - [(c_\beta ^2{\lambda _2} + s_\beta ^2{\lambda _3}){v_2} - {s_\beta }{c_\beta }({\lambda _4} + {\lambda _5}){v_1}]{O_{2i}} - (s_\beta ^2{\kappa _1} + c_\beta ^2{\kappa _2} ){v_s}{O_{3i}},
\\
{g_{ijk}} &=&  - \frac{1}{2}({\lambda _1}{v_1}{O_{1i}} + {\lambda _3}{v_2}{O_{2i}} + {\kappa _1}{v_s}{O_{3i}}){O_{1j}}{O_{1k}}
\nonumber\\
&& - \frac{1}{2}({\lambda _2}{v_2}{O_{2i}} + {\lambda _3}{v_1}{O_{1i}} + {\kappa _2}{v_s}{O_{3i}}){O_{2j}}{O_{2k}}\quad
\nonumber\\
&&  - \frac{{{\lambda _4} + {\lambda _5}}}{2}({v_2}{O_{1i}} + {v_1}{O_{2i}}) {O_{1j}}{O_{2k}} - \frac{1}{2}{\lambda _S}{v_s}{O_{3i}}{O_{3j}}{O_{3k}}
 - \frac{1}{2}{\kappa _1}{v_1} {O_{1i}}{O_{3j}}{O_{3k}}
\nonumber\\
&&  - \frac{1}{2}{\kappa _2}{v_2} {O_{2i}}{O_{3j}}{O_{3k}}
 - \frac{1}{2}{\kappa _1}{v_s}{O_{3i}}{O_{1j}}{O_{1k}} - \frac{1}{2}{\kappa _2}{v_s}{O_{3i}}{O_{2j}}{O_{2k}}.
\end{eqnarray}

By expanding the Lagrangian~\eqref{eq:L_kin}, we obtain the gauge trilinear couplings for the scalars,
\begin{eqnarray}
{\mathcal{L}_{{\mathrm{gauge}}}} &=& \sum\limits_{i = 1}^3 {\left( {{g_{{h_i}WW}}\,{h_i}{W^{ - ,\mu }}W_\mu ^ +  + \frac{1}{2}{g_{{h_i}ZZ}}\,{h_i}{Z^\mu }{Z_\mu } + i{g_{Za{h_i}}}\,{Z_\mu }ai\overleftrightarrow {{\partial ^\mu }}{h_i}} \right)} \
\nonumber\\
&&~  + \sum\limits_{i = 1}^3 {\left( {{g_{{W^ \pm }{H^ \mp }{h_i}}}\,W_\mu ^ + {H^ - }i\overleftrightarrow {{\partial ^\mu }}{h_i} + i\frac{g}{2}\,W_\mu ^ + {H^ - }i\overleftrightarrow {{\partial ^\mu }}a + \mathrm{H.c.}} \right)}
\nonumber\\
&&~ + e\,{A_\mu }{H^ - }i\overleftrightarrow {{\partial ^\mu }}{H^ + }
+ \frac{{g(c_{\mathrm{W}}^2 - s_{\mathrm{W}}^2)}}{{2{c_{\mathrm{W}}}}}\,{Z_\mu }{H^ - }i\overleftrightarrow {{\partial ^\mu }}{H^ + },
\end{eqnarray}
where $s_\mathrm{W} \equiv \sin\theta_\mathrm{W}$.
The derivative symbol $\overleftrightarrow {\partial ^\mu }$ is defined as $F\overleftrightarrow {\partial ^\mu }G = F\partial^\mu G - G\partial^\mu F$.
The coupling coefficients are given by
\begin{eqnarray}
{g_{{h_i}WW}} &=& g{m_W}({c_\beta }{O_{1i}} + {s_\beta }{O_{2i}}),\quad {g_{{h_i}ZZ}} = \frac{{g{m_Z}}}{{{c_{\mathrm{W}}}}}({c_\beta }{O_{1i}} + {s_\beta }{O_{2i}}),
\label{eq:g_hiVV}\\
{g_{Za{h_i}}} &=& \frac{g}{{2{c_{\mathrm{W}}}}}( - {s_\beta }{O_{1i}} + {c_\beta }{O_{2i}}),\quad {g_{{W^ \pm }{H^ \mp }{h_i}}} = \frac{g}{2}( - {s_\beta }{O_{1i}} + {c_\beta }{O_{2i}}).
\end{eqnarray}
 
\section{BSM decay widths of the SM-like Higgs boson}
\label{app:width}

This Appendix gives the decay widths of the SM-like Higgs boson into two-body BSM final states when they are kinematically allowed.
Assuming the SM-like Higgs boson is $h_\mathrm{SM} = h_i$, its invisible decay width at tree level is
\begin{equation}
\Gamma ({h_i} \to \chi \chi ) = \frac{{g_{{h_i}{\chi ^2}}^2}}{{32\pi {m_{{h_i}}}}}\sqrt {1 - \frac{{4m_\chi ^2}}{{m_{{h_i}}^2}}} .
\end{equation}
Moreover, its decay widths into $aa$ and $H^+ H^-$ are  given by
\begin{equation}
\Gamma ({h_i} \to aa) = \frac{{g_{{h_i}{a^2}}^2}}{{32\pi {m_{{h_i}}}}}\sqrt {1 - \frac{{4m_a^2}}{{m_{{h_i}}^2}}} ,\quad \Gamma ({h_i} \to {H^ + }{H^ - }) = \frac{{g_{{h_i}{H^ - }{H^ + }}^2}}{{16\pi {m_{{h_i}}}}}\sqrt {1 - \frac{{4m_{{H^ + }}^2}}{{m_{{h_i}}^2}}} .
\end{equation}

Furthermore, the $h_i \to aZ$ decay width can be expressed as
\begin{equation}
\Gamma ({h_i} \to aZ) = \frac{{g_{Za{h_i}}^2m_{{h_i}}^3}}{{16\pi m_Z^2}}{\lambda ^{3/2}}(1,m_a^2/m_{{h_i}}^2,m_Z^2/m_{{h_i}}^2),
\end{equation}
where the $\lambda$ function is defined by
\begin{equation}
\lambda (x,y,z) \equiv {x^2} + {y^2} + {z^2} - 2xy - 2xz - 2yz.
\end{equation}
The decay width of $h_i \to H^+ W^-$ is given by
\begin{equation}
\Gamma ({h_i} \to {H^+ }{W^-}) = \frac{{g_{{W^ \pm }{H^ \mp }{h_i}}^2m_{{h_i}}^3}}{{16\pi m_W^2}}{\lambda ^{3/2}}(1,m_{{H^ + }}^2/m_{{h_i}}^2,m_W^2/m_{{h_i}}^2),
\end{equation}
which is equal to the decay width of $h_i \to H^- W^+$ .

If $h_\mathrm{SM} = h_2$ or $h_3$, it is possible to decay into $h_1 h_1$ and $h_2 h_2$, whose widths can be commonly expressed as
\begin{equation}
\Gamma ({h_i} \to {h_j}{h_j}) = \frac{{\tilde g_{ijj}^2}}{{8\pi {m_{{h_i}}}}}\sqrt {1 - \frac{{4m_{{h_j}}^2}}{{m_{{h_i}}^2}}}
\end{equation}
with ${{\tilde g}_{ijj}} = {g_{ijj}} + {g_{jij}} + {g_{jji}}$.
If $h_\mathrm{SM} = h_3$, there is another possible decay channel into $h_1 h_2$.
The corresponding width is
\begin{equation}
\Gamma ({h_3} \to {h_1}{h_2}) = \frac{{\tilde g_{123}^2}}{{16\pi {m_{{h_3}}}}}{\lambda ^{1/2}}(1,m_{{h_1}}^2/m_{{h_3}}^2,m_{{h_2}}^2/m_{{h_3}}^2),
\end{equation}
where ${{\tilde g}_{123}} = {g_{123}} + {g_{231}} + {g_{312}} + {g_{213}} + {g_{132}} + {g_{321}}$.

\bibliographystyle{utphys}
\bibliography{ref}

\providecommand{\href}[2]{#2}\begingroup\raggedright\begin{thebibliography}{100}

\bibitem{Bertone:2004pz}
G.~Bertone, D.~Hooper, and J.~Silk, ``{Particle dark matter: Evidence,
  candidates and constraints},''
  \href{http://dx.doi.org/10.1016/j.physrep.2004.08.031}{{\em Phys. Rept.}
  {\bfseries 405} (2005) 279--390},
\href{http://arxiv.org/abs/hep-ph/0404175}{{\ttfamily arXiv:hep-ph/0404175
  [hep-ph]}}.

\bibitem{Feng:2010gw}
J.~L. Feng, ``{Dark Matter Candidates from Particle Physics and Methods of
  Detection},''
  \href{http://dx.doi.org/10.1146/annurev-astro-082708-101659}{{\em Ann. Rev.
  Astron. Astrophys.} {\bfseries 48} (2010) 495--545},
\href{http://arxiv.org/abs/1003.0904}{{\ttfamily arXiv:1003.0904
  [astro-ph.CO]}}.

\bibitem{Young:2016ala}
B.-L. Young, ``{A survey of dark matter and related topics in cosmology},''
  \href{http://dx.doi.org/10.1007/s11467-017-0680-z,
  10.1007/s11467-016-0583-4}{{\em Front. Phys.(Beijing)} {\bfseries 12} (2017)
  121201}.
[Erratum: Front. Phys.(Beijing)12,no.2,121202(2017)].

\bibitem{Akerib:2016vxi}
{\bfseries LUX} Collaboration, D.~S. Akerib {\em et~al.}, ``{Results from a
  search for dark matter in the complete LUX exposure},''
  \href{http://dx.doi.org/10.1103/PhysRevLett.118.021303}{{\em Phys. Rev.
  Lett.} {\bfseries 118} (2017) 021303},
\href{http://arxiv.org/abs/1608.07648}{{\ttfamily arXiv:1608.07648
  [astro-ph.CO]}}.

\bibitem{Cui:2017nnn}
{\bfseries PandaX-II} Collaboration, X.~Cui {\em et~al.}, ``{Dark Matter
  Results From 54-Ton-Day Exposure of PandaX-II Experiment},''
  \href{http://dx.doi.org/10.1103/PhysRevLett.119.181302}{{\em Phys. Rev.
  Lett.} {\bfseries 119} (2017) 181302},
\href{http://arxiv.org/abs/1708.06917}{{\ttfamily arXiv:1708.06917
  [astro-ph.CO]}}.

\bibitem{Aprile:2018dbl}
{\bfseries XENON} Collaboration, E.~Aprile {\em et~al.}, ``{Dark Matter Search
  Results from a One Ton-Year Exposure of XENON1T},''
  \href{http://dx.doi.org/10.1103/PhysRevLett.121.111302}{{\em Phys. Rev.
  Lett.} {\bfseries 121} (2018) 111302},
\href{http://arxiv.org/abs/1805.12562}{{\ttfamily arXiv:1805.12562
  [astro-ph.CO]}}.

\bibitem{Cheung:2012qy}
C.~Cheung, L.~J. Hall, D.~Pinner, and J.~T. Ruderman, ``{Prospects and Blind
  Spots for Neutralino Dark Matter},''
  \href{http://dx.doi.org/10.1007/JHEP05(2013)100}{{\em JHEP} {\bfseries 05}
  (2013) 100},
\href{http://arxiv.org/abs/1211.4873}{{\ttfamily arXiv:1211.4873 [hep-ph]}}.

\bibitem{Banerjee:2016hsk}
S.~Banerjee, S.~Matsumoto, K.~Mukaida, and Y.-L.~S. Tsai, ``{WIMP Dark Matter
  in a Well-Tempered Regime: A case study on Singlet-Doublets Fermionic
  WIMP},'' \href{http://dx.doi.org/10.1007/JHEP11(2016)070}{{\em JHEP}
  {\bfseries 11} (2016) 070},
\href{http://arxiv.org/abs/1603.07387}{{\ttfamily arXiv:1603.07387 [hep-ph]}}.

\bibitem{Cai:2017wdu}
C.~Cai, Z.-H. Yu, and H.-H. Zhang, ``{CEPC Precision of Electroweak Oblique
  Parameters and Weakly Interacting Dark Matter: the Scalar Case},''
  \href{http://dx.doi.org/10.1016/j.nuclphysb.2017.09.007}{{\em Nucl. Phys.}
  {\bfseries B924} (2017) 128--152},
\href{http://arxiv.org/abs/1705.07921}{{\ttfamily arXiv:1705.07921 [hep-ph]}}.

\bibitem{Han:2018gej}
T.~Han, H.~Liu, S.~Mukhopadhyay, and X.~Wang, ``{Dark Matter Blind Spots at
  One-Loop},'' \href{http://dx.doi.org/10.1007/JHEP03(2019)080}{{\em JHEP}
  {\bfseries 03} (2019) 080},
\href{http://arxiv.org/abs/1810.04679}{{\ttfamily arXiv:1810.04679 [hep-ph]}}.

\bibitem{Altmannshofer:2019wjb}
W.~Altmannshofer, B.~Maddock, and S.~Profumo, ``{Doubly Blind Spots in Scalar
  Dark Matter Models},''
\href{http://arxiv.org/abs/1907.01726}{{\ttfamily arXiv:1907.01726 [hep-ph]}}.

\bibitem{Dedes:2014hga}
A.~Dedes and D.~Karamitros, ``{Doublet-Triplet Fermionic Dark Matter},''
  \href{http://dx.doi.org/10.1103/PhysRevD.89.115002}{{\em Phys. Rev.}
  {\bfseries D89} (2014) 115002},
\href{http://arxiv.org/abs/1403.7744}{{\ttfamily arXiv:1403.7744 [hep-ph]}}.

\bibitem{Tait:2016qbg}
T.~M.~P. Tait and Z.-H. Yu, ``{Triplet-Quadruplet Dark Matter},''
  \href{http://dx.doi.org/10.1007/JHEP03(2016)204}{{\em JHEP} {\bfseries 03}
  (2016) 204},
\href{http://arxiv.org/abs/1601.01354}{{\ttfamily arXiv:1601.01354 [hep-ph]}}.

\bibitem{Arcadi:2016kmk}
G.~Arcadi, C.~Gross, O.~Lebedev, Y.~Mambrini, S.~Pokorski, and T.~Toma,
  ``{Multicomponent Dark Matter from Gauge Symmetry},''
  \href{http://dx.doi.org/10.1007/JHEP12(2016)081}{{\em JHEP} {\bfseries 12}
  (2016) 081},
\href{http://arxiv.org/abs/1611.00365}{{\ttfamily arXiv:1611.00365 [hep-ph]}}.

\bibitem{Cai:2016sjz}
C.~Cai, Z.-H. Yu, and H.-H. Zhang, ``{CEPC Precision of Electroweak Oblique
  Parameters and Weakly Interacting Dark Matter: the Fermionic Case},''
  \href{http://dx.doi.org/10.1016/j.nuclphysb.2017.05.015}{{\em Nucl. Phys.}
  {\bfseries B921} (2017) 181--210},
\href{http://arxiv.org/abs/1611.02186}{{\ttfamily arXiv:1611.02186 [hep-ph]}}.

\bibitem{Xiang:2017yfs}
Q.-F. Xiang, X.-J. Bi, P.-F. Yin, and Z.-H. Yu, ``{Exploring Fermionic Dark
  Matter via Higgs Boson Precision Measurements at the Circular Electron
  Positron Collider},''
  \href{http://dx.doi.org/10.1103/PhysRevD.97.055004}{{\em Phys. Rev.}
  {\bfseries D97} (2018) 055004},
\href{http://arxiv.org/abs/1707.03094}{{\ttfamily arXiv:1707.03094 [hep-ph]}}.

\bibitem{Wang:2017sxx}
J.-W. Wang, X.-J. Bi, Q.-F. Xiang, P.-F. Yin, and Z.-H. Yu, ``{Exploring
  triplet-quadruplet fermionic dark matter at the LHC and future colliders},''
  \href{http://dx.doi.org/10.1103/PhysRevD.97.035021}{{\em Phys. Rev.}
  {\bfseries D97} (2018) 035021},
\href{http://arxiv.org/abs/1711.05622}{{\ttfamily arXiv:1711.05622 [hep-ph]}}.

\bibitem{Ipek:2014gua}
S.~Ipek, D.~McKeen, and A.~E. Nelson, ``{A Renormalizable Model for the
  Galactic Center Gamma Ray Excess from Dark Matter Annihilation},''
  \href{http://dx.doi.org/10.1103/PhysRevD.90.055021}{{\em Phys. Rev.}
  {\bfseries D90} (2014) 055021},
\href{http://arxiv.org/abs/1404.3716}{{\ttfamily arXiv:1404.3716 [hep-ph]}}.

\bibitem{Berlin:2015wwa}
A.~Berlin, S.~Gori, T.~Lin, and L.-T. Wang, ``{Pseudoscalar Portal Dark
  Matter},'' \href{http://dx.doi.org/10.1103/PhysRevD.92.015005}{{\em Phys.
  Rev.} {\bfseries D92} (2015) 015005},
\href{http://arxiv.org/abs/1502.06000}{{\ttfamily arXiv:1502.06000 [hep-ph]}}.

\bibitem{No:2015xqa}
J.~M. No, ``{Looking through the pseudoscalar portal into dark matter: Novel
  mono-Higgs and mono-Z signatures at the LHC},''
  \href{http://dx.doi.org/10.1103/PhysRevD.93.031701}{{\em Phys. Rev.}
  {\bfseries D93} (2016) 031701},
\href{http://arxiv.org/abs/1509.01110}{{\ttfamily arXiv:1509.01110 [hep-ph]}}.

\bibitem{Goncalves:2016iyg}
D.~Goncalves, P.~A.~N. Machado, and J.~M. No, ``{Simplified Models for Dark
  Matter Face their Consistent Completions},''
  \href{http://dx.doi.org/10.1103/PhysRevD.95.055027}{{\em Phys. Rev.}
  {\bfseries D95} (2017) 055027},
\href{http://arxiv.org/abs/1611.04593}{{\ttfamily arXiv:1611.04593 [hep-ph]}}.

\bibitem{Haisch:2016gry}
U.~Haisch, P.~Pani, and G.~Polesello, ``{Determining the CP nature of spin-0
  mediators in associated production of dark matter and $ t\overline{t} $
  pairs},'' \href{http://dx.doi.org/10.1007/JHEP02(2017)131}{{\em JHEP}
  {\bfseries 02} (2017) 131},
\href{http://arxiv.org/abs/1611.09841}{{\ttfamily arXiv:1611.09841 [hep-ph]}}.

\bibitem{Bauer:2017ota}
M.~Bauer, U.~Haisch, and F.~Kahlhoefer, ``{Simplified dark matter models with
  two Higgs doublets: I. Pseudoscalar mediators},''
  \href{http://dx.doi.org/10.1007/JHEP05(2017)138}{{\em JHEP} {\bfseries 05}
  (2017) 138},
\href{http://arxiv.org/abs/1701.07427}{{\ttfamily arXiv:1701.07427 [hep-ph]}}.

\bibitem{Tunney:2017yfp}
P.~Tunney, J.~M. No, and M.~Fairbairn, ``{Probing the pseudoscalar portal to
  dark matter via $\bar bbZ(\to\ell\ell)+ \not{E}_T$ : From the LHC to the
  Galactic Center excess},''
  \href{http://dx.doi.org/10.1103/PhysRevD.96.095020}{{\em Phys. Rev.}
  {\bfseries D96} (2017) 095020},
\href{http://arxiv.org/abs/1705.09670}{{\ttfamily arXiv:1705.09670 [hep-ph]}}.

\bibitem{Barducci:2016fue}
D.~Barducci, A.~Bharucha, N.~Desai, M.~Frigerio, B.~Fuks, A.~Goudelis,
  S.~Kulkarni, G.~Polesello, and D.~Sengupta, ``{Monojet searches for
  momentum-dependent dark matter interactions},''
  \href{http://dx.doi.org/10.1007/JHEP01(2017)078}{{\em JHEP} {\bfseries 01}
  (2017) 078},
\href{http://arxiv.org/abs/1609.07490}{{\ttfamily arXiv:1609.07490 [hep-ph]}}.

\bibitem{Gross:2017dan}
C.~Gross, O.~Lebedev, and T.~Toma, ``{Cancellation Mechanism for
  Dark-Matter–Nucleon Interaction},''
  \href{http://dx.doi.org/10.1103/PhysRevLett.119.191801}{{\em Phys. Rev.
  Lett.} {\bfseries 119} (2017) 191801},
\href{http://arxiv.org/abs/1708.02253}{{\ttfamily arXiv:1708.02253 [hep-ph]}}.

\bibitem{Balkin:2018tma}
R.~Balkin, M.~Ruhdorfer, E.~Salvioni, and A.~Weiler, ``{Dark matter shifts away
  from direct detection},''
  \href{http://dx.doi.org/10.1088/1475-7516/2018/11/050}{{\em JCAP} {\bfseries
  1811} (2018) 050},
\href{http://arxiv.org/abs/1809.09106}{{\ttfamily arXiv:1809.09106 [hep-ph]}}.

\bibitem{Huitu:2018gbc}
K.~Huitu, N.~Koivunen, O.~Lebedev, S.~Mondal, and T.~Toma, ``{Probing
  pseudo-Goldstone dark matter at the LHC},''
\href{http://arxiv.org/abs/1812.05952}{{\ttfamily arXiv:1812.05952 [hep-ph]}}.

\bibitem{Alanne:2018zjm}
T.~Alanne, M.~Heikinheimo, V.~Keus, N.~Koivunen, and K.~Tuominen, ``{Direct and
  indirect probes of Goldstone dark matter},''
  \href{http://dx.doi.org/10.1103/PhysRevD.99.075028}{{\em Phys. Rev.}
  {\bfseries D99} (2019) 075028},
\href{http://arxiv.org/abs/1812.05996}{{\ttfamily arXiv:1812.05996 [hep-ph]}}.

\bibitem{Kannike:2019wsn}
K.~Kannike and M.~Raidal, ``{Phase Transitions and Gravitational Wave Tests of
  Pseudo-Goldstone Dark Matter in the Softly Broken U(1) Scalar Singlet
  Model},'' \href{http://dx.doi.org/10.1103/PhysRevD.99.115010}{{\em Phys.
  Rev.} {\bfseries D99} (2019) 115010},
\href{http://arxiv.org/abs/1901.03333}{{\ttfamily arXiv:1901.03333 [hep-ph]}}.

\bibitem{Karamitros:2019ewv}
D.~Karamitros, ``{Pseudo Nambu-Goldstone Dark Matter: Examples of Vanishing
  Direct Detection Cross Section},''
  \href{http://dx.doi.org/10.1103/PhysRevD.99.095036}{{\em Phys. Rev.}
  {\bfseries D99} (2019) 095036},
\href{http://arxiv.org/abs/1901.09751}{{\ttfamily arXiv:1901.09751 [hep-ph]}}.

\bibitem{Cline:2019okt}
J.~M. Cline and T.~Toma, ``{Pseudo-Goldstone dark matter confronts cosmic ray
  and collider anomalies},''
\href{http://arxiv.org/abs/1906.02175}{{\ttfamily arXiv:1906.02175 [hep-ph]}}.

\bibitem{Azevedo:2018exj}
D.~Azevedo, M.~Duch, B.~Grzadkowski, D.~Huang, M.~Iglicki, and R.~Santos,
  ``{One-loop contribution to dark-matter-nucleon scattering in the
  pseudo-scalar dark matter model},''
  \href{http://dx.doi.org/10.1007/JHEP01(2019)138}{{\em JHEP} {\bfseries 01}
  (2019) 138},
\href{http://arxiv.org/abs/1810.06105}{{\ttfamily arXiv:1810.06105 [hep-ph]}}.

\bibitem{Ishiwata:2018sdi}
K.~Ishiwata and T.~Toma, ``{Probing pseudo Nambu-Goldstone boson dark matter at
  loop level},'' \href{http://dx.doi.org/10.1007/JHEP12(2018)089}{{\em JHEP}
  {\bfseries 12} (2018) 089},
\href{http://arxiv.org/abs/1810.08139}{{\ttfamily arXiv:1810.08139 [hep-ph]}}.

\bibitem{Branco:2011iw}
G.~C. Branco, P.~M. Ferreira, L.~Lavoura, M.~N. Rebelo, M.~Sher, and J.~P.
  Silva, ``{Theory and phenomenology of two-Higgs-doublet models},''
  \href{http://dx.doi.org/10.1016/j.physrep.2012.02.002}{{\em Phys. Rept.}
  {\bfseries 516} (2012) 1--102},
\href{http://arxiv.org/abs/1106.0034}{{\ttfamily arXiv:1106.0034 [hep-ph]}}.

\bibitem{Haber:1984rc}
H.~E. Haber and G.~L. Kane, ``{The Search for Supersymmetry: Probing Physics
  Beyond the Standard Model},''
\href{http://dx.doi.org/10.1016/0370-1573(85)90051-1}{{\em Phys. Rept.}
  {\bfseries 117} (1985) 75--263}.

\bibitem{Kim:1986ax}
J.~E. Kim, ``{Light Pseudoscalars, Particle Physics and Cosmology},''
\href{http://dx.doi.org/10.1016/0370-1573(87)90017-2}{{\em Phys. Rept.}
  {\bfseries 150} (1987) 1--177}.

\bibitem{Turok:1990zg}
N.~Turok and J.~Zadrozny, ``{Electroweak baryogenesis in the two doublet
  model},''
\href{http://dx.doi.org/10.1016/0550-3213(91)90356-3}{{\em Nucl. Phys.}
  {\bfseries B358} (1991) 471--493}.

\bibitem{Ko:2015fxa}
P.~Ko, Y.~Omura, and C.~Yu, ``{Higgs and dark matter physics in the type-II
  two-Higgs-doublet model inspired by $E_{6}$ GUT},''
  \href{http://dx.doi.org/10.1007/JHEP06(2015)034}{{\em JHEP} {\bfseries 06}
  (2015) 034},
\href{http://arxiv.org/abs/1502.00262}{{\ttfamily arXiv:1502.00262 [hep-ph]}}.

\bibitem{Bell:2016ekl}
N.~F. Bell, G.~Busoni, and I.~W. Sanderson, ``{Self-consistent Dark Matter
  Simplified Models with an s-channel scalar mediator},''
  \href{http://dx.doi.org/10.1088/1475-7516/2017/03/015}{{\em JCAP} {\bfseries
  1703} (2017) 015},
\href{http://arxiv.org/abs/1612.03475}{{\ttfamily arXiv:1612.03475 [hep-ph]}}.

\bibitem{Chang:2017gla}
C.-F. Chang, X.-G. He, and J.~Tandean, ``{Two-Higgs-Doublet-Portal Dark-Matter
  Models in Light of Direct Search and LHC Data},''
  \href{http://dx.doi.org/10.1007/JHEP04(2017)107}{{\em JHEP} {\bfseries 04}
  (2017) 107},
\href{http://arxiv.org/abs/1702.02924}{{\ttfamily arXiv:1702.02924 [hep-ph]}}.

\bibitem{Bell:2017rgi}
N.~F. Bell, G.~Busoni, and I.~W. Sanderson, ``{Two Higgs Doublet Dark Matter
  Portal},'' \href{http://dx.doi.org/10.1088/1475-7516/2018/01/015}{{\em JCAP}
  {\bfseries 1801} (2018) 015},
\href{http://arxiv.org/abs/1710.10764}{{\ttfamily arXiv:1710.10764 [hep-ph]}}.

\bibitem{Dey:2019lyr}
A.~Dey, J.~Lahiri, and B.~Mukhopadhyaya, ``{LHC signals of a heavy doublet
  Higgs as dark matter portal: cut-based approach and improvement with gradient
  boosting and neural networks},''
\href{http://arxiv.org/abs/1905.02242}{{\ttfamily arXiv:1905.02242 [hep-ph]}}.

\bibitem{Glashow:1976nt}
S.~L. Glashow and S.~Weinberg, ``{Natural Conservation Laws for Neutral
  Currents},''
\href{http://dx.doi.org/10.1103/PhysRevD.15.1958}{{\em Phys. Rev.} {\bfseries
  D15} (1977) 1958}.

\bibitem{Paschos:1976ay}
E.~A. Paschos, ``{Diagonal Neutral Currents},''
\href{http://dx.doi.org/10.1103/PhysRevD.15.1966}{{\em Phys. Rev.} {\bfseries
  D15} (1977) 1966}.

\bibitem{Camargo:2019ukv}
D.~A. Camargo, M.~D. Campos, T.~B. de~Melo, and F.~S. Queiroz, ``{A Two Higgs
  Doublet Model for Dark Matter and Neutrino Masses},''
  \href{http://dx.doi.org/10.1016/j.physletb.2019.06.020}{{\em Phys. Lett.}
  {\bfseries B795} (2019) 319--326},
\href{http://arxiv.org/abs/1901.05476}{{\ttfamily arXiv:1901.05476 [hep-ph]}}.

\bibitem{Gunion:2002zf}
J.~F. Gunion and H.~E. Haber, ``{The CP conserving two Higgs doublet model: The
  Approach to the decoupling limit},''
  \href{http://dx.doi.org/10.1103/PhysRevD.67.075019}{{\em Phys. Rev.}
  {\bfseries D67} (2003) 075019},
\href{http://arxiv.org/abs/hep-ph/0207010}{{\ttfamily arXiv:hep-ph/0207010
  [hep-ph]}}.

\bibitem{Carena:2013ooa}
M.~Carena, I.~Low, N.~R. Shah, and C.~E.~M. Wagner, ``{Impersonating the
  Standard Model Higgs Boson: Alignment without Decoupling},''
  \href{http://dx.doi.org/10.1007/JHEP04(2014)015}{{\em JHEP} {\bfseries 04}
  (2014) 015},
\href{http://arxiv.org/abs/1310.2248}{{\ttfamily arXiv:1310.2248 [hep-ph]}}.

\bibitem{Dev:2014yca}
P.~S. Bhupal~Dev and A.~Pilaftsis, ``{Maximally Symmetric Two Higgs Doublet
  Model with Natural Standard Model Alignment},''
  \href{http://dx.doi.org/10.1007/JHEP11(2015)147,
  10.1007/JHEP12(2014)024}{{\em JHEP} {\bfseries 12} (2014) 024},
  \href{http://arxiv.org/abs/1408.3405}{{\ttfamily arXiv:1408.3405 [hep-ph]}}.
[Erratum: JHEP11,147(2015)].

\bibitem{Georgi:1978ri}
H.~Georgi and D.~V. Nanopoulos, ``{Suppression of Flavor Changing Effects From
  Neutral Spinless Meson Exchange in Gauge Theories},''
\href{http://dx.doi.org/10.1016/0370-2693(79)90433-7}{{\em Phys. Lett.}
  {\bfseries 82B} (1979) 95--96}.

\bibitem{Donoghue:1978cj}
J.~F. Donoghue and L.~F. Li, ``{Properties of Charged Higgs Bosons},''
\href{http://dx.doi.org/10.1103/PhysRevD.19.945}{{\em Phys. Rev.} {\bfseries
  D19} (1979) 945}.

\bibitem{Tanabashi:2018oca}
{\bfseries Particle Data Group} Collaboration, M.~Tanabashi {\em et~al.},
  ``{Review of Particle Physics},''
\href{http://dx.doi.org/10.1103/PhysRevD.98.030001}{{\em Phys. Rev.} {\bfseries
  D98} (2018) 030001}.

\bibitem{Heinemeyer:2013tqa}
{\bfseries LHC Higgs Cross Section Working Group} Collaboration, J.~R. Andersen
  {\em et~al.}, ``{Handbook of LHC Higgs Cross Sections: 3. Higgs
  Properties},''
\href{http://arxiv.org/abs/1307.1347}{{\ttfamily arXiv:1307.1347 [hep-ph]}}.

\bibitem{Bernon:2015hsa}
J.~Bernon and B.~Dumont, ``{Lilith: a tool for constraining new physics from
  Higgs measurements},''
  \href{http://dx.doi.org/10.1140/epjc/s10052-015-3645-9}{{\em Eur. Phys. J.}
  {\bfseries C75} (2015) 440},
\href{http://arxiv.org/abs/1502.04138}{{\ttfamily arXiv:1502.04138 [hep-ph]}}.

\bibitem{Aaltonen:2013ioz}
{\bfseries CDF, D0} Collaboration, T.~Aaltonen {\em et~al.}, ``{Higgs Boson
  Studies at the Tevatron},''
  \href{http://dx.doi.org/10.1103/PhysRevD.88.052014}{{\em Phys. Rev.}
  {\bfseries D88} (2013) 052014},
\href{http://arxiv.org/abs/1303.6346}{{\ttfamily arXiv:1303.6346 [hep-ex]}}.

\bibitem{Aad:2014iia}
{\bfseries ATLAS} Collaboration, G.~Aad {\em et~al.}, ``{Search for Invisible
  Decays of a Higgs Boson Produced in Association with a Z Boson in ATLAS},''
  \href{http://dx.doi.org/10.1103/PhysRevLett.112.201802}{{\em Phys. Rev.
  Lett.} {\bfseries 112} (2014) 201802},
\href{http://arxiv.org/abs/1402.3244}{{\ttfamily arXiv:1402.3244 [hep-ex]}}.

\bibitem{Aad:2014eha}
{\bfseries ATLAS} Collaboration, G.~Aad {\em et~al.}, ``{Measurement of Higgs
  boson production in the diphoton decay channel in pp collisions at
  center-of-mass energies of 7 and 8 TeV with the ATLAS detector},''
  \href{http://dx.doi.org/10.1103/PhysRevD.90.112015}{{\em Phys. Rev.}
  {\bfseries D90} (2014) 112015},
\href{http://arxiv.org/abs/1408.7084}{{\ttfamily arXiv:1408.7084 [hep-ex]}}.

\bibitem{Aad:2014xzb}
{\bfseries ATLAS} Collaboration, G.~Aad {\em et~al.}, ``{Search for the
  $b\bar{b}$ decay of the Standard Model Higgs boson in associated $(W/Z)H$
  production with the ATLAS detector},''
  \href{http://dx.doi.org/10.1007/JHEP01(2015)069}{{\em JHEP} {\bfseries 01}
  (2015) 069},
\href{http://arxiv.org/abs/1409.6212}{{\ttfamily arXiv:1409.6212 [hep-ex]}}.

\bibitem{ATLAS:2015yda}
{\bfseries ATLAS} Collaboration, G.~Aad {\em et~al.}, ``{Search for an
  Invisibly Decaying Higgs Boson Produced via Vector Boson Fusion in $pp$
  Collisions at $\sqrt{s}=8$ TeV using the ATLAS Detector at the LHC},''
\href{http://cds.cern.ch/record/2002121}{ATLAS-CONF-2015-004}.

\bibitem{Aad:2015gra}
{\bfseries ATLAS} Collaboration, G.~Aad {\em et~al.}, ``{Search for the
  Standard Model Higgs boson produced in association with top quarks and
  decaying into $b\bar{b}$ in pp collisions at $\sqrt{s}$ = 8 TeV with the
  ATLAS detector},''
  \href{http://dx.doi.org/10.1140/epjc/s10052-015-3543-1}{{\em Eur. Phys. J.}
  {\bfseries C75} (2015) 349},
\href{http://arxiv.org/abs/1503.05066}{{\ttfamily arXiv:1503.05066 [hep-ex]}}.

\bibitem{Aad:2015iha}
{\bfseries ATLAS} Collaboration, G.~Aad {\em et~al.}, ``{Search for the
  associated production of the Higgs boson with a top quark pair in multilepton
  final states with the ATLAS detector},''
  \href{http://dx.doi.org/10.1016/j.physletb.2015.07.079}{{\em Phys. Lett.}
  {\bfseries B749} (2015) 519--541},
\href{http://arxiv.org/abs/1506.05988}{{\ttfamily arXiv:1506.05988 [hep-ex]}}.

\bibitem{Aad:2015ona}
{\bfseries ATLAS} Collaboration, G.~Aad {\em et~al.}, ``{Study of (W/Z)H
  production and Higgs boson couplings using $H \rightarrow WW^{\ast}$ decays
  with the ATLAS detector},''
  \href{http://dx.doi.org/10.1007/JHEP08(2015)137}{{\em JHEP} {\bfseries 08}
  (2015) 137},
\href{http://arxiv.org/abs/1506.06641}{{\ttfamily arXiv:1506.06641 [hep-ex]}}.

\bibitem{Aad:2015gba}
{\bfseries ATLAS} Collaboration, G.~Aad {\em et~al.}, ``{Measurements of the
  Higgs boson production and decay rates and coupling strengths using pp
  collision data at $\sqrt{s}=7$ and 8 TeV in the ATLAS experiment},''
  \href{http://dx.doi.org/10.1140/epjc/s10052-015-3769-y}{{\em Eur. Phys. J.}
  {\bfseries C76} (2016) 6},
\href{http://arxiv.org/abs/1507.04548}{{\ttfamily arXiv:1507.04548 [hep-ex]}}.

\bibitem{Chatrchyan:2014tja}
{\bfseries CMS} Collaboration, S.~Chatrchyan {\em et~al.}, ``{Search for
  invisible decays of Higgs bosons in the vector boson fusion and associated ZH
  production modes},''
  \href{http://dx.doi.org/10.1140/epjc/s10052-014-2980-6}{{\em Eur. Phys. J.}
  {\bfseries C74} (2014) 2980},
\href{http://arxiv.org/abs/1404.1344}{{\ttfamily arXiv:1404.1344 [hep-ex]}}.

\bibitem{Khachatryan:2014qaa}
{\bfseries CMS} Collaboration, V.~Khachatryan {\em et~al.}, ``{Search for the
  associated production of the Higgs boson with a top-quark pair},''
  \href{http://dx.doi.org/10.1007/JHEP09(2014)087,
  10.1007/JHEP10(2014)106}{{\em JHEP} {\bfseries 09} (2014) 087},
  \href{http://arxiv.org/abs/1408.1682}{{\ttfamily arXiv:1408.1682 [hep-ex]}}.
[Erratum: JHEP10,106(2014)].

\bibitem{Khachatryan:2014jba}
{\bfseries CMS} Collaboration, V.~Khachatryan {\em et~al.}, ``{Precise
  determination of the mass of the Higgs boson and tests of compatibility of
  its couplings with the standard model predictions using proton collisions at
  7 and 8 $\,\text {TeV}$},''
  \href{http://dx.doi.org/10.1140/epjc/s10052-015-3351-7}{{\em Eur. Phys. J.}
  {\bfseries C75} (2015) 212},
\href{http://arxiv.org/abs/1412.8662}{{\ttfamily arXiv:1412.8662 [hep-ex]}}.

\bibitem{Khachatryan:2015ila}
{\bfseries CMS} Collaboration, V.~Khachatryan {\em et~al.}, ``{Search for a
  Standard Model Higgs Boson Produced in Association with a Top-Quark Pair and
  Decaying to Bottom Quarks Using a Matrix Element Method},''
  \href{http://dx.doi.org/10.1140/epjc/s10052-015-3454-1}{{\em Eur. Phys. J.}
  {\bfseries C75} (2015) 251},
\href{http://arxiv.org/abs/1502.02485}{{\ttfamily arXiv:1502.02485 [hep-ex]}}.

\bibitem{Khachatryan:2015bnx}
{\bfseries CMS} Collaboration, V.~Khachatryan {\em et~al.}, ``{Search for the
  standard model Higgs boson produced through vector boson fusion and decaying
  to $b \overline{b}$},''
  \href{http://dx.doi.org/10.1103/PhysRevD.92.032008}{{\em Phys. Rev.}
  {\bfseries D92} (2015) 032008},
\href{http://arxiv.org/abs/1506.01010}{{\ttfamily arXiv:1506.01010 [hep-ex]}}.

\bibitem{ATLAS:2016bza}
{\bfseries ATLAS} Collaboration, G.~Aad {\em et~al.}, ``{Search for new
  phenomena in the $Z(\rightarrow\ell\ell) + E_{\mathrm{T}}^{\mathrm{miss}}$
  final state at $\sqrt{s}$ = 13 TeV with the ATLAS detector},''
\href{http://cds.cern.ch/record/2206138}{ATLAS-CONF-2016-056}.

\bibitem{ATLAS:2016ldo}
{\bfseries ATLAS} Collaboration, G.~Aad {\em et~al.}, ``{Search for the
  Associated Production of a Higgs Boson and a Top Quark Pair in Multilepton
  Final States with the ATLAS Detector},''
\href{http://cds.cern.ch/record/2206153}{ATLAS-CONF-2016-058}.

\bibitem{ATLAS:2016lgh}
{\bfseries ATLAS} Collaboration, G.~Aad {\em et~al.}, ``{Search for Higgs boson
  production via weak boson fusion and decaying to $b \bar b$ in association
  with a high-energy photon in the ATLAS detector},''
\href{http://cds.cern.ch/record/2206201}{ATLAS-CONF-2016-063}.

\bibitem{ATLAS:2016nke}
{\bfseries ATLAS} Collaboration, G.~Aad {\em et~al.}, ``{Measurement of
  fiducial, differential and production cross sections in the
  $H\to\gamma\gamma$ decay channel with 13.3 fb$^{-1}$ of 13 TeV proton-proton
  collision data with the ATLAS detector},''
\href{http://cds.cern.ch/record/2206210}{ATLAS-CONF-2016-067}.

\bibitem{ATLAS:2016oum}
{\bfseries ATLAS} Collaboration, G.~Aad {\em et~al.}, ``{Study of the Higgs
  boson properties and search for high-mass scalar resonances in the $H
  \rightarrow ZZ^* \rightarrow 4\ell$ decay channel at $\sqrt{s}$ = 13 TeV with
  the ATLAS detector},''
\href{http://cds.cern.ch/record/2206253}{ATLAS-CONF-2016-079}.

\bibitem{ATLAS:2016awy}
{\bfseries ATLAS} Collaboration, G.~Aad {\em et~al.}, ``{Search for the
  Standard Model Higgs boson produced in association with top quarks and
  decaying into $b\overline{b}$ in $pp$ collisions at $\sqrt{s}$ = 13 TeV with
  the ATLAS detector},''
\href{http://cds.cern.ch/record/2206255}{ATLAS-CONF-2016-080}.

\bibitem{ATLAS:2016pkl}
{\bfseries ATLAS} Collaboration, G.~Aad {\em et~al.}, ``{Search for the
  Standard Model Higgs boson produced in association with a vector boson and
  decaying to a $b\bar{b}$ pair in $pp$ collisions at 13 TeV using the ATLAS
  detector},''
\href{http://cds.cern.ch/record/2206813}{ATLAS-CONF-2016-091}.

\bibitem{ATLAS:2016gld}
{\bfseries ATLAS} Collaboration, G.~Aad {\em et~al.}, ``{Measurements of the
  Higgs boson production cross section via Vector Boson Fusion and associated
  $WH$ production in the $WW^{\ast} \to \ell\nu\ell\nu$ decay mode with the
  ATLAS detector at $\sqrt{s}$ = 13 TeV},''
\href{http://cds.cern.ch/record/2231811}{ATLAS-CONF-2016-112}.

\bibitem{ATLAS:2017syx}
{\bfseries ATLAS} Collaboration, G.~Aad {\em et~al.}, ``{Search for the dimuon
  decay of the Higgs boson in $pp$ collisions at $\sqrt{s}$ = 13 TeV with the
  ATLAS detector},''
\href{http://cds.cern.ch/record/2257726}{ATLAS-CONF-2017-014}.

\bibitem{CMS:2016nfx}
{\bfseries CMS} Collaboration, V.~Khachatryan {\em et~al.}, ``{First results on
  Higgs to WW at $\sqrt{s}=13~\mathrm{TeV}$},''
\href{http://cds.cern.ch/record/2161793}{CMS-PAS-HIG-15-003}.

\bibitem{CMS:2016mmc}
{\bfseries CMS} Collaboration, V.~Khachatryan {\em et~al.}, ``{VBF H to bb
  using the 2015 data sample},''
\href{http://cds.cern.ch/record/2160154}{CMS-PAS-HIG-16-003}.

\bibitem{CMS:2016jjx}
{\bfseries CMS} Collaboration, V.~Khachatryan {\em et~al.}, ``{Search for
  invisible decays of a Higgs boson produced via vector boson fusion at
  $\sqrt{s}=13$ TeV.},''
\href{http://cds.cern.ch/record/2142460}{CMS-PAS-HIG-16-009}.

\bibitem{CMS:2016ixj}
{\bfseries CMS} Collaboration, V.~Khachatryan {\em et~al.}, ``{Updated
  measurements of Higgs boson production in the diphoton decay channel at
  $\sqrt{s}=13~\textrm{TeV}$ in pp collisions at CMS},''
\href{http://cds.cern.ch/record/2205275}{CMS-PAS-HIG-16-020}.

\bibitem{CMS:2016zbb}
{\bfseries CMS} Collaboration, V.~Khachatryan {\em et~al.}, ``{Search for
  $\mathrm{t\overline{t}H}$ production in the $\mathrm{H}\rightarrow
  \mathrm{b\overline{b}}$ decay channel with 2016 pp collision data at
  $\sqrt{s}=13~\mathrm{TeV}$},''
\href{http://cds.cern.ch/record/2231510}{CMS-PAS-HIG-16-038}.

\bibitem{CMS:2017jkd}
{\bfseries CMS} Collaboration, V.~Khachatryan {\em et~al.}, ``{Measurements of
  properties of the Higgs boson decaying into four leptons in pp collisions at
  sqrt{s} = 13 TeV},''
\href{http://cds.cern.ch/record/2256357}{CMS-PAS-HIG-16-041}.

\bibitem{CMS:2017lgc}
{\bfseries CMS} Collaboration, V.~Khachatryan {\em et~al.}, ``{Search for the
  associated production of a Higgs boson with a top quark pair in final states
  with a $\tau$ lepton at $\sqrt{s} = 13~\mathrm{TeV}$},''
\href{http://cds.cern.ch/record/2257067}{CMS-PAS-HIG-17-003}.

\bibitem{Alloul:2013bka}
A.~Alloul, N.~D. Christensen, C.~Degrande, C.~Duhr, and B.~Fuks, ``{FeynRules
  2.0 - A complete toolbox for tree-level phenomenology},''
  \href{http://dx.doi.org/10.1016/j.cpc.2014.04.012}{{\em Comput. Phys.
  Commun.} {\bfseries 185} (2014) 2250--2300},
\href{http://arxiv.org/abs/1310.1921}{{\ttfamily arXiv:1310.1921 [hep-ph]}}.

\bibitem{Alwall:2014hca}
J.~Alwall, R.~Frederix, S.~Frixione, V.~Hirschi, F.~Maltoni, O.~Mattelaer,
  H.~S. Shao, T.~Stelzer, P.~Torrielli, and M.~Zaro, ``{The automated
  computation of tree-level and next-to-leading order differential cross
  sections, and their matching to parton shower simulations},''
  \href{http://dx.doi.org/10.1007/JHEP07(2014)079}{{\em JHEP} {\bfseries 07}
  (2014) 079},
\href{http://arxiv.org/abs/1405.0301}{{\ttfamily arXiv:1405.0301 [hep-ph]}}.

\bibitem{Ambrogi:2018jqj}
F.~Ambrogi, C.~Arina, M.~Backovic, J.~Heisig, F.~Maltoni, L.~Mantani,
  O.~Mattelaer, and G.~Mohlabeng, ``{MadDM v.3.0: a Comprehensive Tool for Dark
  Matter Studies},'' \href{http://dx.doi.org/10.1016/j.dark.2018.11.009}{{\em
  Phys. Dark Univ.} {\bfseries 24} (2019) 100249},
\href{http://arxiv.org/abs/1804.00044}{{\ttfamily arXiv:1804.00044 [hep-ph]}}.

\bibitem{Ade:2015xua}
{\bfseries Planck} Collaboration, P.~A.~R. Ade {\em et~al.}, ``{Planck 2015
  results. XIII. Cosmological parameters},''
  \href{http://dx.doi.org/10.1051/0004-6361/201525830}{{\em Astron. Astrophys.}
  {\bfseries 594} (2016) A13},
\href{http://arxiv.org/abs/1502.01589}{{\ttfamily arXiv:1502.01589
  [astro-ph.CO]}}.

\bibitem{Lin:2000qq}
W.~B. Lin, D.~H. Huang, X.~Zhang, and R.~H. Brandenberger, ``{Nonthermal
  production of WIMPs and the subgalactic structure of the universe},''
  \href{http://dx.doi.org/10.1103/PhysRevLett.86.954}{{\em Phys. Rev. Lett.}
  {\bfseries 86} (2001) 954},
\href{http://arxiv.org/abs/astro-ph/0009003}{{\ttfamily arXiv:astro-ph/0009003
  [astro-ph]}}.

\bibitem{Fujii:2002kr}
M.~Fujii and K.~Hamaguchi, ``{Nonthermal dark matter via Affleck-Dine
  baryogenesis and its detection possibility},''
  \href{http://dx.doi.org/10.1103/PhysRevD.66.083501}{{\em Phys. Rev.}
  {\bfseries D66} (2002) 083501},
\href{http://arxiv.org/abs/hep-ph/0205044}{{\ttfamily arXiv:hep-ph/0205044
  [hep-ph]}}.

\bibitem{Springel:2008cc}
V.~Springel, J.~Wang, M.~Vogelsberger, A.~Ludlow, A.~Jenkins, A.~Helmi, J.~F.
  Navarro, C.~S. Frenk, and S.~D.~M. White, ``{The Aquarius Project: the
  subhalos of galactic halos},''
  \href{http://dx.doi.org/10.1111/j.1365-2966.2008.14066.x}{{\em Mon. Not. Roy.
  Astron. Soc.} {\bfseries 391} (2008) 1685--1711},
\href{http://arxiv.org/abs/0809.0898}{{\ttfamily arXiv:0809.0898 [astro-ph]}}.

\bibitem{Diemand:2008in}
J.~Diemand, M.~Kuhlen, P.~Madau, M.~Zemp, B.~Moore, D.~Potter, and J.~Stadel,
  ``{Clumps and streams in the local dark matter distribution},''
  \href{http://dx.doi.org/10.1038/nature07153}{{\em Nature} {\bfseries 454}
  (2008) 735--738},
\href{http://arxiv.org/abs/0805.1244}{{\ttfamily arXiv:0805.1244 [astro-ph]}}.

\bibitem{Strigari:2013iaa}
L.~E. Strigari, ``{Galactic Searches for Dark Matter},''
  \href{http://dx.doi.org/10.1016/j.physrep.2013.05.004}{{\em Phys. Rept.}
  {\bfseries 531} (2013) 1--88},
\href{http://arxiv.org/abs/1211.7090}{{\ttfamily arXiv:1211.7090
  [astro-ph.CO]}}.

\bibitem{Mateo:1998wg}
M.~Mateo, ``{Dwarf galaxies of the Local Group},''
  \href{http://dx.doi.org/10.1146/annurev.astro.36.1.435}{{\em Ann. Rev.
  Astron. Astrophys.} {\bfseries 36} (1998) 435--506},
\href{http://arxiv.org/abs/astro-ph/9810070}{{\ttfamily arXiv:astro-ph/9810070
  [astro-ph]}}.

\bibitem{Gallagher:2003nx}
J.~S. Gallagher, G.~J. Madsen, R.~J. Reynolds, E.~K. Grebel, and T.~A.
  Smecker-Hane, ``{A search for ionized gas in the Draco and Ursa Minor dwarf
  spheroidal galaxies},'' \href{http://dx.doi.org/10.1086/373951}{{\em
  Astrophys. J.} {\bfseries 588} (2003) 326--330},
\href{http://arxiv.org/abs/astro-ph/0301228}{{\ttfamily arXiv:astro-ph/0301228
  [astro-ph]}}.

\bibitem{Grcevich:2009gt}
J.~Grcevich and M.~E. Putman, ``{HI in Local Group Dwarf Galaxies and Stripping
  by the Galactic Halo},'' \href{http://dx.doi.org/10.1088/0004-637X/721/1/922,
  10.1088/0004-637X/696/1/385}{{\em Astrophys. J.} {\bfseries 696} (2009)
  385--395}, \href{http://arxiv.org/abs/0901.4975}{{\ttfamily arXiv:0901.4975
  [astro-ph.GA]}}.
[Erratum: Astrophys. J.721,922(2010)].

\bibitem{Walker:2009zp}
M.~G. Walker, M.~Mateo, E.~W. Olszewski, J.~Penarrubia, N.~W. Evans, and
  G.~Gilmore, ``{A Universal Mass Profile for Dwarf Spheroidal Galaxies},''
  \href{http://dx.doi.org/10.1088/0004-637X/704/2/1274,
  10.1088/0004-637X/710/1/886}{{\em Astrophys. J.} {\bfseries 704} (2009)
  1274--1287}, \href{http://arxiv.org/abs/0906.0341}{{\ttfamily arXiv:0906.0341
  [astro-ph.CO]}}.
[Erratum: Astrophys. J.710,886(2010)].

\bibitem{Gondolo:1990dk}
P.~Gondolo and G.~Gelmini, ``{Cosmic abundances of stable particles: Improved
  analysis},''
\href{http://dx.doi.org/10.1016/0550-3213(91)90438-4}{{\em Nucl. Phys.}
  {\bfseries B360} (1991) 145--179}.

\bibitem{Griest:1990kh}
K.~Griest and D.~Seckel, ``{Three exceptions in the calculation of relic
  abundances},''
\href{http://dx.doi.org/10.1103/PhysRevD.43.3191}{{\em Phys. Rev.} {\bfseries
  D43} (1991) 3191--3203}.

\bibitem{Ahnen:2016qkx}
{\bfseries MAGIC, Fermi-LAT} Collaboration, M.~L. Ahnen {\em et~al.}, ``{Limits
  to Dark Matter Annihilation Cross-Section from a Combined Analysis of MAGIC
  and Fermi-LAT Observations of Dwarf Satellite Galaxies},''
  \href{http://dx.doi.org/10.1088/1475-7516/2016/02/039}{{\em JCAP} {\bfseries
  1602} (2016) 039},
\href{http://arxiv.org/abs/1601.06590}{{\ttfamily arXiv:1601.06590
  [astro-ph.HE]}}.

\bibitem{Navarro:1996gj}
J.~F. Navarro, C.~S. Frenk, and S.~D.~M. White, ``{A Universal density profile
  from hierarchical clustering},'' \href{http://dx.doi.org/10.1086/304888}{{\em
  Astrophys. J.} {\bfseries 490} (1997) 493--508},
\href{http://arxiv.org/abs/astro-ph/9611107}{{\ttfamily arXiv:astro-ph/9611107
  [astro-ph]}}.

\bibitem{CEPCStudyGroup:2018ghi}
{\bfseries CEPC Study Group} Collaboration, M.~Dong {\em et~al.}, ``{CEPC
  Conceptual Design Report: Volume 2 - Physics \& Detector},''
\href{http://arxiv.org/abs/1811.10545}{{\ttfamily arXiv:1811.10545 [hep-ex]}}.

\bibitem{Baer:2013cma}
H.~Baer, T.~Barklow, K.~Fujii, Y.~Gao, A.~Hoang, S.~Kanemura, J.~List, H.~E.
  Logan, A.~Nomerotski, M.~Perelstein, {\em et~al.}, ``{The International
  Linear Collider Technical Design Report - Volume 2: Physics},''
\href{http://arxiv.org/abs/1306.6352}{{\ttfamily arXiv:1306.6352 [hep-ph]}}.

\bibitem{Abada:2019zxq}
{\bfseries FCC} Collaboration, A.~Abada {\em et~al.}, ``{FCC-ee: The Lepton
  Collider},''
\href{http://dx.doi.org/10.1140/epjst/e2019-900045-4}{{\em Eur. Phys. J. ST}
  {\bfseries 228} (2019) 261--623}.

\end{thebibliography}\endgroup

\end{document}